\documentclass[groupedaddress, aps, longbibliography, reprint, prx]{revtex4-1}

\usepackage{graphicx}
\usepackage{bm}
\usepackage{bbold}
\usepackage{physics}
\usepackage{xcolor}
\usepackage{enumitem}
\usepackage{amsmath, amssymb}
\usepackage[normalem]{ulem}
\usepackage{natbib}
\usepackage{physics}
\usepackage{comment}
\usepackage{wasysym}
\usepackage{stmaryrd}

\usepackage[colorlinks=true]{hyperref}
\hypersetup{citecolor = blue, linkcolor=blue, urlcolor=blue}

\newcommand{\occupied}{\ensuremath{\mathbin{\newmoon}}} 
\newcommand{\vacuum}{\ensuremath{\mathbin{\fullmoon}}}
\newcommand{\free}{\ensuremath{\mathbin{\oplus}}}

\newcommand{\shadownorm}[1]{\ensuremath{\| #1 \|_{\rm sh}}}

\begin{document}

\title{Operator relaxation and the optimal depth of classical shadows}

\author{Matteo Ippoliti}
\author{Yaodong Li}
\author{Tibor Rakovszky}
\author{Vedika Khemani}
\affiliation{Department of Physics, Stanford University, Stanford, CA 94305, USA}

\begin{abstract}
Classical shadows are a powerful method for learning many properties of quantum states in a sample-efficient manner, by making use of randomized measurements. Here we study the sample complexity of learning the expectation value of Pauli operators via ``shallow shadows'', a recently-proposed version of classical shadows in which the randomization step is effected by a local unitary circuit of variable depth $t$. 
We show that the shadow norm (the quantity controlling the sample complexity) is expressed in terms of properties of the Heisenberg time evolution of operators under the randomizing (``twirling'') circuit---namely the evolution of the {\it weight distribution} characterizing the number of sites on which an operator acts nontrivially.
For spatially-contiguous Pauli operators of weight $k$, this entails a competition between two processes: {\it operator spreading} (whereby the support of an operator grows over time, increasing its weight) and {\it operator relaxation} (whereby the bulk of the operator develops an equilibrium density of identity operators, decreasing its weight). 
From this simple picture we derive 
(i) an upper bound on the shadow norm which, for depth $t\sim \log(k)$, guarantees an exponential gain in sample complexity over the $t=0$ protocol in any spatial dimension, and 
(ii) quantitative results in one dimension within a mean-field approximation, including a universal subleading correction to the optimal depth, found to be in excellent agreement with infinite matrix product state numerical simulations.
Our work connects fundamental ideas in quantum many-body dynamics to applications in quantum information science, and paves the way to highly-optimized protocols for learning different properties of quantum states.
\end{abstract}

\maketitle


{\it Introduction.}
The development of controllable quantum simulators has enabled the creation of complex and highly entangled quantum states in laboratory settings, leading to exciting new developments in quantum information science and many-body physics ~\cite{arute_quantum_2019, altman_quantum_2021, mi_information_2021, wu_strong_2021, egan_fault-tolerant_2021, acharya_suppressing_2022, ebadi_quantum_2021, semeghini_probing_2021}. These advances raise the issue of how to { efficiently} characterize such quantum states. 
Full quantum state tomography requires exponentially many measurements in the size of the system~\cite{haah_sample-optimal_2017}, motivating the need for more scalable and efficient state-learning protocols. Recent progress in this direction has come from the development of {\it classical shadows}~\cite{aaronson_shadow_2018, huang_predicting_2020, paini_approximate_2019, chen_robust_2021, acharya_informationally_2021, struchalin_experimental_2021, levy_classical_2021, zhao_fermionic_2021, wan_matchgate_2022, huang_provably_2022, bu_classical_2022, kunjummen_shadow_2022, shivam_classical_2022}, a method to extract many physical properties of states with a dramatically smaller number of measurements.
In this work, we shed light on the inner workings of classical shadows by making connections to foundational ideas in quantum dynamics on the spreading and equilibration of operators.

\begin{figure}
    \centering
    \includegraphics[width=\columnwidth]{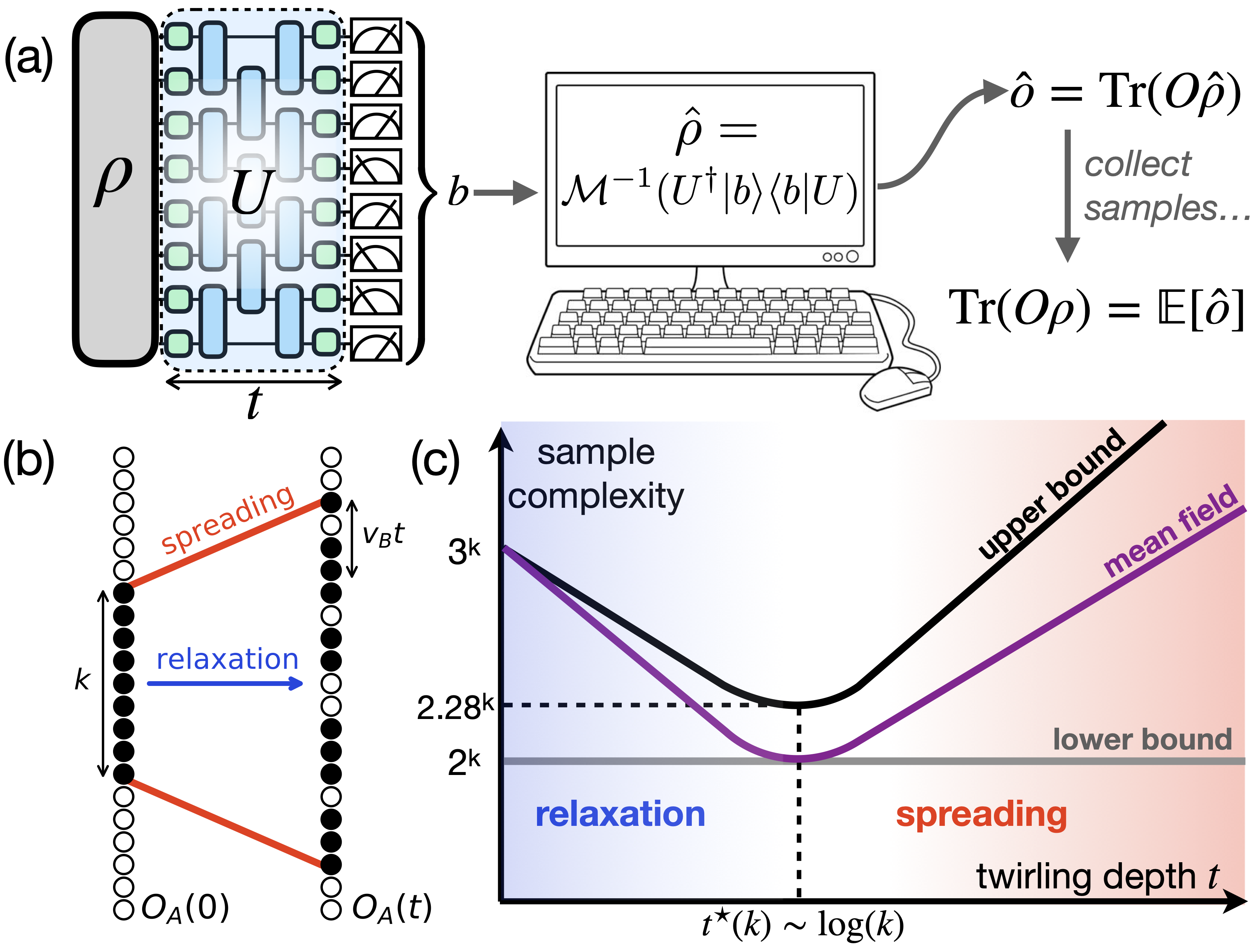}
    \caption{ 
    (a) Schematic of classical shadows via shallow circuits: a state $\rho$ is randomized by a ``twirling'' circuit $U$ of depth $t$, then measured; data is classically processed to estimate Pauli expectation values.
    (b) Operator spreading and relaxation under chaotic dynamics. $\vacuum/\occupied$ denote identity and traceless Pauli matrices, respectively. 
    (c) Summary of main results of this work. The competition between operator spreading and relaxation determines the optimal sample complexity of learning Pauli expectation values.}
    \label{fig:schematic}
\end{figure}

Classical shadows use {\it randomized measurements}~\cite{elben_statistical_2019, brydges_probing_2019, elben_randomized_2022} to form a compact representation of a many-body quantum state, Fig.~\ref{fig:schematic}(a).
The state $\rho$ is first transformed by a random unitary operation $U$ (chosen from a suitable ``twirling ensemble''), then projectively measured, yielding a computational basis state $\ket{b}$. The measured basis state is then rotated backwards (on a classical computer), giving a ``snapshot" $\hat{\sigma}_{U,b} = U^\dagger \ketbra{b}{b} U$. 
The average of these snapshots (over twirling unitaries and measurement outcomes) is related to the true state $\rho$ by a quantum channel, $\mathbb{E}_{U,b}[\hat{\sigma}_{U,b}] = \mathcal{M}(\rho)$.  If the measurements are {\it tomographically complete}~\cite{huang_predicting_2020}, the channel $\mathcal{M}$ can be inverted (again on a classical computer) to produce ``inverted snapshots'' $\hat{\rho}_{U,b} = \mathcal{M}^{-1}(\hat{\sigma}_{U,b})$. These form a compact, approximate description of the quantum state $\rho$---its {\it classical shadow}~\cite{huang_predicting_2020}.
From this description one can extract many properties of the state, which remarkably do not have to be specified in advance---the general philosophy of the method is to {\it ``measure first, ask questions later''}~\cite{elben_randomized_2022}. 

The usefulness of classical shadows depends on their {\it sample complexity}, i.e., the number of experimental samples needed in order to estimate a certain property of $\rho$ within a given error. 
To learn an expectation value ${\rm Tr}(\rho O)$, one builds estimators $\hat{o}_{U,b} = {\rm Tr}(\hat{\rho}_{U,b}O)$ that yield the desired value in expectation ($\mathbb{E}_{U,b}[\hat{o}_{U,b}] = {\rm Tr}(\rho O)$). The sample complexity is determined by the variance of $\hat o$, captured by the {\it shadow norm} $\shadownorm{O}$, itself a function of the twirling ensemble. 
The freedom in choosing the twirling ensemble can thus be leveraged to optimize the learnability of certain properties of a quantum state.
For instance, ``local twirling'' (where $U = \bigotimes_i u_i$ is a product of single-qubit random unitaries) gives $\shadownorm{O}^2 = 3^k$ for Pauli operators, where $k$ is the number of qubits on which $O$ acts nontrivially; 
this is best suited to learning the value of few-body operators. 
On the opposite end, ``global twirling'' (where $U$ is a random Clifford unitary on the whole Hilbert space) gives $\shadownorm{O}^2 = {\rm Tr}(O^\dagger O)$, which favors learning e.g. the fidelity with a pure many-body state $O = \ketbra{\psi}{\psi}$, but performs poorly on Pauli operators ($\shadownorm{O}^2 = 2^N$) irrespective of locality~\cite{huang_predicting_2020}.

Intermediate schemes, dubbed {\it shallow shadows}, have been recently proposed~\cite{akhtar_scalable_2022, bertoni_shallow_2022, arienzo_closed-form_2022} and use twirling ensembles made of shallow quantum circuits, whose depth $t$ can be tuned to interpolate between the local and global twirling limits. 
The finite depth $t$ makes these easier to implement on quantum hardware, and enables efficient classical computation of $\hat{\sigma}$ and $\hat{\rho}$ via tensor-network methods~\cite{akhtar_scalable_2022, bertoni_shallow_2022}. 
Surprisingly, these schemes were numerically observed to perform better than local twirling for estimating the expectation value of contiguous, multi-site Pauli operators (interesting examples of such operators include string order parameters for characterizing topological phases~\cite{kennedy_hidden_1992, haegeman_order_2012} and check operators of a quantum code~\cite{gottesman_stabilizer_1997}). 
The optimal depth $t^\star(k)$ for a Pauli operator acting on $k$ contiguous sites was observed numerically to scale as $\text{polylog}(k)$ in one dimension~\cite{akhtar_scalable_2022}, with a significant gain in sample complexity over the local twlirling protocol. The physical mechanism behind this behavior has remained elusive thus far. 

Here we analyze this problem analytically and find a mapping of the shadow norm to the dynamics of {\it Hamming weight} (the number of sites on which a Pauli operator acts nontrivially, henceforth just `weight') under the twirling evolution. 
This mapping reveals that the optimal depth for the estimation of contiguous Pauli operators is determined by the competition of two processes under chaotic unitary dynamics, 
sketched in Fig.~\ref{fig:schematic}(b): {\it operator spreading}~\cite{maldacena_bound_2016, swingle_measuring_2016, nahum_operator_2018, khemani_operator_2018, von_keyserlingk_operator_2018, rakovszky_diffusive_2018} and {\it operator relaxation}, to be defined below. 
Based on this picture, we prove that at depth $t^{\star}(k) \sim \log(k)$, shallow shadows realize an exponential-in-$k$ gain in sample complexity over local twirling in any finite spatial dimension. 
We further develop an analytical mean-field approximation for the shadow norm in one dimension, indicating that at depth $t^\star(k)$ the sample complexity nearly saturates a lower bound ($\sim 2^k$, up to $\text{poly}(k)$ corrections), as sketched in Fig.~\ref{fig:schematic}(c); 
the prediction shows excellent agreement with numerics on large Pauli operators (up to $k=1000$) in infinite 1D systems.

Our results shed light on the inner workings of the classical shadows protocol and how it relates to fundamental aspects of quantum dynamics. At the same time, they give a practical, operational meaning to ideas about 
operator dynamics, and promise applications towards highly optimized classical shadows protocols for near-term quantum devices.

{\it Shadow norm and operator weight.} 
We begin by deriving a relationship between the shadow norm and operator dynamics valid if the twirling ensemble is {\it locally scrambled}~\cite{kuo_markovian_2020, hu_classical_2021}, i.e., such that measure ${\rm d}U$ over the ensemble is invariant under $U \mapsto VU$ and $U \mapsto UV$ for all product Clifford unitaries~\cite{gottesman_heisenberg_1998} $V = \bigotimes_i v_i$, $v_i \in {\sf Cliff}(q)$ (this holds for local and global twirling, as well as for shallow shadows~\cite{akhtar_scalable_2022, bertoni_shallow_2022, arienzo_closed-form_2022}).

We will consider a system of $q$-state qudits arranged on a $d$-dimensional lattice consisting of $N$ qudits. For qudits with $q>2$, we use ``generalized Pauli operators'' defined by products of clock and shift unitary operators~\cite{gheorghiu_standard_2014}. The measurement channel reads
\begin{equation}
    \mathcal{M}(\rho) = \sum_{b} \int{\rm d}U 
    \overbrace{\bra{b} U \rho U^\dagger \ket{b}}^{{\sf Prob}(b|\rho,U)} \overbrace{U^\dagger \ketbra{b}{b} U}^{\text{snapshot } \hat{\sigma}_{U,b}},
\end{equation}
where $b$ ranges over all $D = q^N$ computational basis states.

All Pauli operators are eigenmodes of the channel~\cite{bu_classical_2022, akhtar_scalable_2022, bertoni_shallow_2022}, and the eigenvalue depends solely on the twirling ensemble and on the support $A$ of the Pauli operator: $\mathcal{M}[O_A] = \lambda_A O_A$, where $O_A$ denotes a Pauli operator supported in region $A$.
The eigenvalues can be expressed as~\footnote{See Supplementary Material for the derivation of Eq.~\eqref{eq:shdn_mainresult} and Eq.~\eqref{eq:opt_depth}, results on Brownian circuits, computation of velocity scales, estimation of non-contiguous operators, and additional details on the random walk mapping, mean-field approximation and numerical methods.}
\begin{equation}
    \lambda_A = \sum_{w = 1}^N \pi_{A,t}(w) (q+1)^{-w},
    \label{eq:lambda_vs_weight}
\end{equation}
where $\pi_{A,t}(w)$ is the averaged {\it weight distribution}~\cite{qi_measuring_2019} of the twirled operator $O_A(t) \equiv UO_A U^\dagger$:
\begin{equation}
    \pi_{A,t}(w) = \sum_{P:\ |P|=w} \mathbb{E}_U \left| D^{-1} {\rm Tr}(PO_A(t)) \right|^2.
    \label{eq:weight_dist}
\end{equation}
The sum runs over Pauli operators $P$, and $|P|$ is the weight 
of $P$.

With this result, we can exactly compute the shadow norm:
$\shadownorm{O_A}^2 = {\rm Tr}(O_A^\dagger \mathcal{M}^{-1}[O_A])/D = \lambda_A^{-1}$~\cite{akhtar_scalable_2022, bertoni_shallow_2022, Note1}.
Combined with 
Eq.~\eqref{eq:lambda_vs_weight}, this yields an exact relationship between the shadow norm and the weight distribution of a twirled operator,
\begin{align}
    \shadownorm{O_A}^2 & = \left[ \overline{(q+1)^{-w}} \right]^{-1} \label{eq:shdn_mainresult}
\end{align}
where the overline denotes averaging over $w$ according to $\pi_{A,t}(w)$.
Eq.~\eqref{eq:shdn_mainresult} constitutes one of the main results of our work. 

Eq.~\eqref{eq:shdn_mainresult} reproduces the well-known results for local and global twirling of qubits ($3^k$ and $2^N$ respectively~\cite{huang_predicting_2020}) in the $t=0$ and $t\to\infty$ limits~\cite{Note1}.
However, our result allows us to understand the behavior of the shadow norm away from these well-know limits, by leveraging the connection to the dynamics of operator weight under chaotic evolution (i.e. the twirling ensemble $U$) as a function of time (i.e. the variable depth $t$). 


{\it Relaxation of operator weight.} 
We focus on Pauli operators whose support $A$ is a spatially-contiguous region (though our results also have implications for more general, non-contiguous Pauli operators~\cite{Note1}). We consider twirling ensembles of {\it diluted} random brickwork circuits, i.e. circuits where each gate is Haar-random~\footnote{The same results would be obtained with any unitary 2-design (e.g. random Clifford gates).} with probability $\epsilon$ and is the identity otherwise. These include conventional random circuits ($\epsilon=1$), but allow us to slow down the twirling dynamics and discretize time more finely. 
To study the dynamics of operator weight during twirling, we introduce ``occupation'' variables $n_i$ ($n_i=0$ if a Pauli operator is the identity at site $i$, $n_i = 1$ otherwise).
Before twirling, we have a {\it fully-packed} Pauli operator in region $A$: $n_i = 1$ iff $i \in A$.
As the twirling depth $t$ increases, two things happen:
{(i) {\it Operator spreading}}---the boundary of the operator moves outwards, so that $\overline{n_i}(t) >0$ also on sites $i\notin A$ that were initially empty, leading to an {\it increase} in weight; and
{(ii) {\it Operator relaxation}}---the bulk of the operator relaxes from its fully-packed initial state ($n_i = 1$ $\forall\ i\in A$) towards an equilibrium density $\overline{n_i} (t) \to 1-q^{-2}$ (when all $q^2$ Pauli operators are equally likely), leading to a {\it decrease} in weight. 

As the latter is a bulk effect, it always dominates (at early times) for a sufficiently large region $A$. Thus the shadow norm must initially decrease from its $t=0$ value (local twirling), before eventually becoming dominated by operator spreading and increasing again towards its $t\to\infty$ value (global twirling), implying a minimum at some finite optimal depth $t^\star$.

\begin{figure}
    \centering
    \includegraphics[width=\columnwidth]{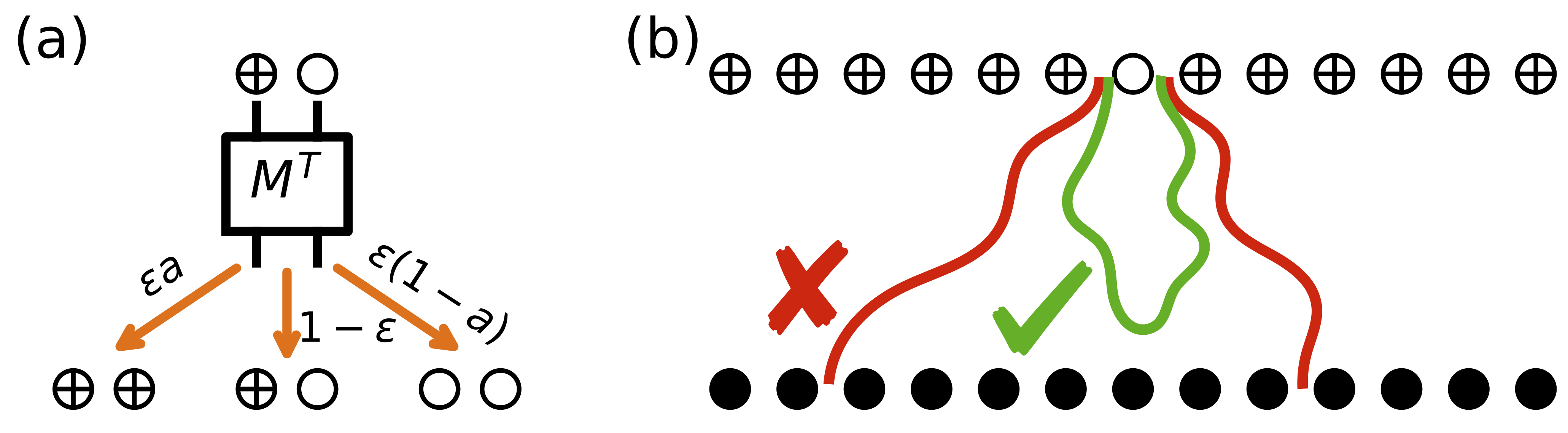}
    \caption{
    (a) Update rules for a domain wall between $\free$ and $\vacuum$ states.
    (b) Random-walk calculation for the average density of holes $\overline{h_i}(t)$: if the two walkers fail to annihilate within $t$ steps, the diagram vanishes.
    }
    \label{fig:rw}
\end{figure}

To characterize the relaxation process, we focus on an infinite, fully-packed Pauli operator, and consider the average occupation of a site $\overline{n_i} (t) $ as a function of twirling depth $t$. 
For the twirling ensembles under consideration 
this problem can be addressed analytically in one spatial dimension. 
We leverage the fact that the vector of occupation probabilities $p_{\mathbf n}$ ($\mathbf{n}\in \{0,1\}^N$ labels occupation configurations) evolves under the circuit-averaged dynamics via a Markov process, $p'_{\mathbf n} = \sum_{\mathbf m} \mathbb{M}_{\mathbf n, \mathbf m} p_{\mathbf m}$ with $\mathbb{M}$ a stochastic matrix ($\sum_{\mathbf m} \mathbb{M}_{\mathbf m, \mathbf{n}} = 1$ $\forall \mathbf{n}$), to solve for the local occupation number analytically~\cite{nahum_operator_2018,von_keyserlingk_operator_2018}. 
We focus on the ``density of holes'' $\overline{h_i}$ (${h}_i \equiv 1-n_i$) and introduce vectors in the binary space of (identity, traceless Pauli): $|\vacuum) = (1,0)^T$, $|\occupied) = (0,1)^T$, and $|\free) = (1,1)^T$. 
The fully-packed initial state $p^\text{init}_{\mathbf n} = \prod_i \delta_{n_i,1}$
evolves under the averaged circuit into a final state $p^\text{final}_{\mathbf n}$, and we have $\overline{h_i} = \sum_{\mathbf n} p^\text{final}_{\mathbf n} \delta_{n_i, 0}$, corresponding to a matrix element 
$ (\!\cdots\! \free\! \free\! \vacuum\! \free\! \free\! \cdots\! | \mathbb{M}_t |\!\cdots\! \occupied\! \occupied\! \occupied\! \cdots\!)$ where $\mathbb{M}_t$ is the transition matrix for the averaged depth-$t$ twirling circuit.

It is advantageous to consider the {\it backward} evolution $\mathbb{M}_t^T$ acting on the state $|\!\cdots\!\free\!\free\!\free\!\vacuum\!\free\!\free\!\free\!\cdots)$:
we have 
$M^T|\!\free\!\vacuum\!) = \epsilon a |\!\free\!\free\!) + (1-\epsilon) |\!\free\!\vacuum\!) + \epsilon (1-a) |\!\vacuum\!\vacuum\!)$ (Fig.~\ref{fig:rw}(a)), where $M$ is the transition matrix for a single two-qudit gate, $a = 1/(q^2+1)$, and $\epsilon$ is the dilution parameter (see \cite{Note1}). 
Moreover we have $M^T|\!\vacuum\!\vacuum\!) = |\!\vacuum\!\vacuum\!)$ (unitary invariance of the identity operator) and $M^T|\!\free\!\free\!) = |\!\free\!\free\!)$ (conservation of total probability under the Markov process~\footnote{Note $(\!\free\!\free\!|p)$ yields the sum of the two-site probability distribution $p_{n_1,n_2}$; conservation of total probability, i.e. $(\!\free\!\free\!|M|p) = (\!\free\!\free\!|p)$ $\forall p$, imposes $(\!\free\!\free\!|M = (\!\free\!\free\!|$.}).
Thus the structure of a domain of $\vacuum$ in a background of $\free$ is preserved under $\mathbb{M}_t^T$, and domain walls undergo a random walk with a bias that tends to expand the $\vacuum$ domain.
When the domain walls are adjacent, they may annihilate, leading to an all-$\free$ state which is invariant under $M^T$ and yields a contribution $(\free|\occupied)^N = 1$; 
if the domain of $\vacuum$ survives all the way to $t=0$, the result vanishes as it involves at least one overlap $(\vacuum|\occupied)=0$ (Fig.~\ref{fig:rw}(b)).

In all, the average density of holes $\overline{h_i}(t)$ equals the probability that the two random walkers annihilate in $t$ steps or less;
conversely, $\overline{n_i}(t)$ equals their {\it survival probability}, which can be computed analytically: at large $t$,
\begin{equation}
    \overline{ n_i} (t) = 1 - q^{-2} + c t^{-3/2} e^{-\gamma t} + \dots \label{eq:n_bulk}
\end{equation}
for any site $i$ in the bulk of the operator, with $c>0$ a constant and $\dots$ denoting subleading corrections in $t$~\cite{Note1}.
The relaxation rate $\gamma$ is related to the circuit's {\it entanglement velocity} $v_E$ (which sets the decay of half-system purity as $\sim q^{-v_E t}$)~\cite{nahum_quantum_2017} via $\gamma = 2\ln(q)v_E$, see \cite{Note1};
the $t^{-3/2}$ is a universal correction related to the first return of a random walker in one dimension~\cite{fisher_walks_1984}. 
We conjecture that the convergence to equilibrium is exponential in any finite spatial dimension, and numerically verify it in two dimensions~\cite{Note1}.

\begin{figure}
    \centering
    \includegraphics[width=\columnwidth]{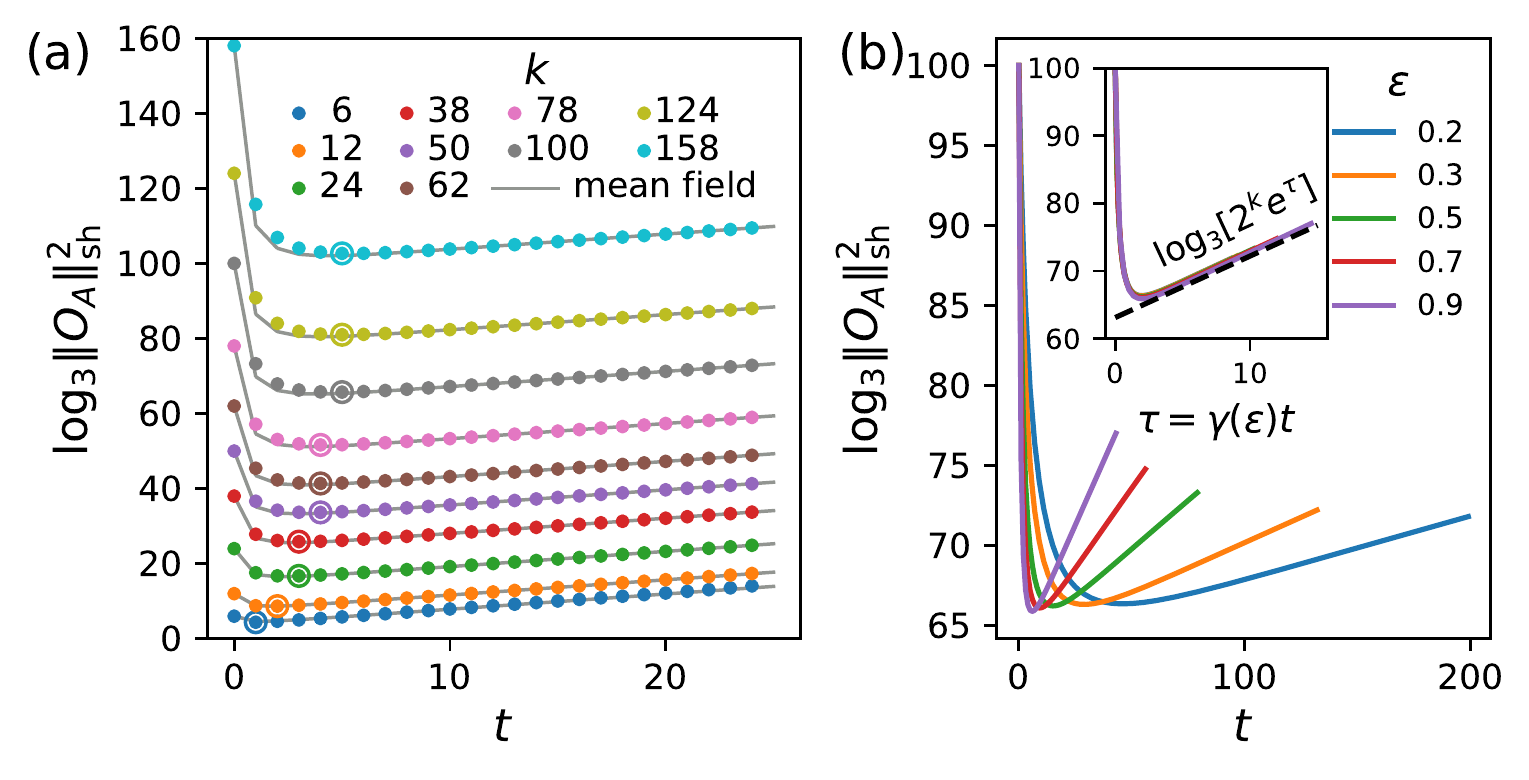}
    \caption{(a) Shadow norm of a weight-$k$ Pauli string $O_A$ in an infinite 1D system of qubits ($q=2$), under twirlig by depth-$t$ brickwork circuits of Haar-random gates (no gate dilution, $\epsilon=1$). Data from iMPS simulations with bond dimension $\chi = 2048$. Circled dots indicate the optimal depth.
    (b) Same quantity for fixed $k=100$ and variable gate dilution $\epsilon$. 
    Inset: same data as a function of ``effective depth'' $\tau = \gamma(\epsilon) t$, compared to $q^k e^{\gamma t}$ (dashed line).
    }
    \label{fig:itebd_shdn}
\end{figure}


{\it Scaling of the optimal depth.} 
With these key results in hand, we return to the question of the optimal depth. 
From Eq.~\eqref{eq:shdn_mainresult} and Jensen's inequality, we have $\shadownorm{O_A}^2 \leq (q+1)^{\overline{w}}$; in one dimension, the average weight obeys $\overline{w}(t) = \sum_i \overline{n_i}(t) \simeq \overline{\ell}(t) \overline{n_\text{bulk}}(t)$, with $\overline{\ell}(t) = k + 2v_Bt$ the average spatial length of the twirled operator, which spreads with butterfly velocity $v_B$~\cite{hosur_chaos_2016, nahum_operator_2018, von_keyserlingk_operator_2018}, and $\overline{n_\text{bulk}}(t)$ the bulk density of traceless Paulis, Eq.~\eqref{eq:n_bulk} (the structure of the operator's fronts can be neglected at large $k$). 
The bound is minimized at depth
\begin{equation}
    t^\star(k) = \gamma^{-1} \left( \ln(k) - \frac{3}{2} \ln\ln(k) + o(\ln\ln (k) ) \right) \label{eq:opt_depth}
\end{equation}
(see \cite{Note1}).
At $t=t^\star(k)$, the shadow norm is bounded above by $(q+1)^{(1-q^{-2})k} \times {\rm poly}(k) $, {\it exponentially smaller} than the $t=0$ (local twirling) value of $(q+1)^k$; e.g., for qubits ($q=2$) the scaling is $3^{\frac{3}{4}k} \simeq 2.28^k$ vs $3^k$. 
The scaling $\log(k)$ (as opposed to more general ${\rm polylog}(k)$~\cite{akhtar_scalable_2022}) is especially important as it ensures an MPO representation for $\mathcal{M}^{-1}$ with ${\rm poly}(k)$ bond dimension, key to the classical computational cost of the method~\cite{akhtar_scalable_2022, bertoni_shallow_2022}.

We conjecture that $t = t^\star(k)$ minimizes not just the upper bound $(q+1)^{\overline{w}}$, but the shadow norm itself, and that the achievable scaling of the latter is ${\rm poly}(k)\times q^k$---nearly saturating the $q^k$ lower bound obtained by full relaxation with no spreading.
This is supported by an analytical calculation within a mean-field approximation, where we neglect correlations between occupations $n_i$, $n_j$ at different sites, see \cite{Note1}. 
We find that $\shadownorm{O_A}$ is dominated by Pauli operators of size $k + 2v_B^{\rm sp}t$, with a renormalized ``saddle-point butterfly velocity'' $v_B^{\rm sp}$ smaller than the original $v_B$, and equal to the entanglement velocity $v_E = \gamma / \ln(q^2)$. This predicts the late-time behavior $\shadownorm{O_A}^2 \sim q^{k+ 2v_B^{\rm sp}t} = q^k e^{\gamma t}$. 
Minimizing the mean-field shadow norm over $t$ yields the same $t^\star(k)$ as in Eq.~\eqref{eq:opt_depth}, and thus the optimal shadow norm $\sim k q^k$.

It follows also that shallow shadows can be advantageous over local twirling not just for operators with contiguous support, but also for various types of non-contiguous operators, notably including typical random Pauli strings on a finite segment~\cite{Note1}. 

\begin{figure}
    \centering
    \includegraphics[width=\columnwidth]{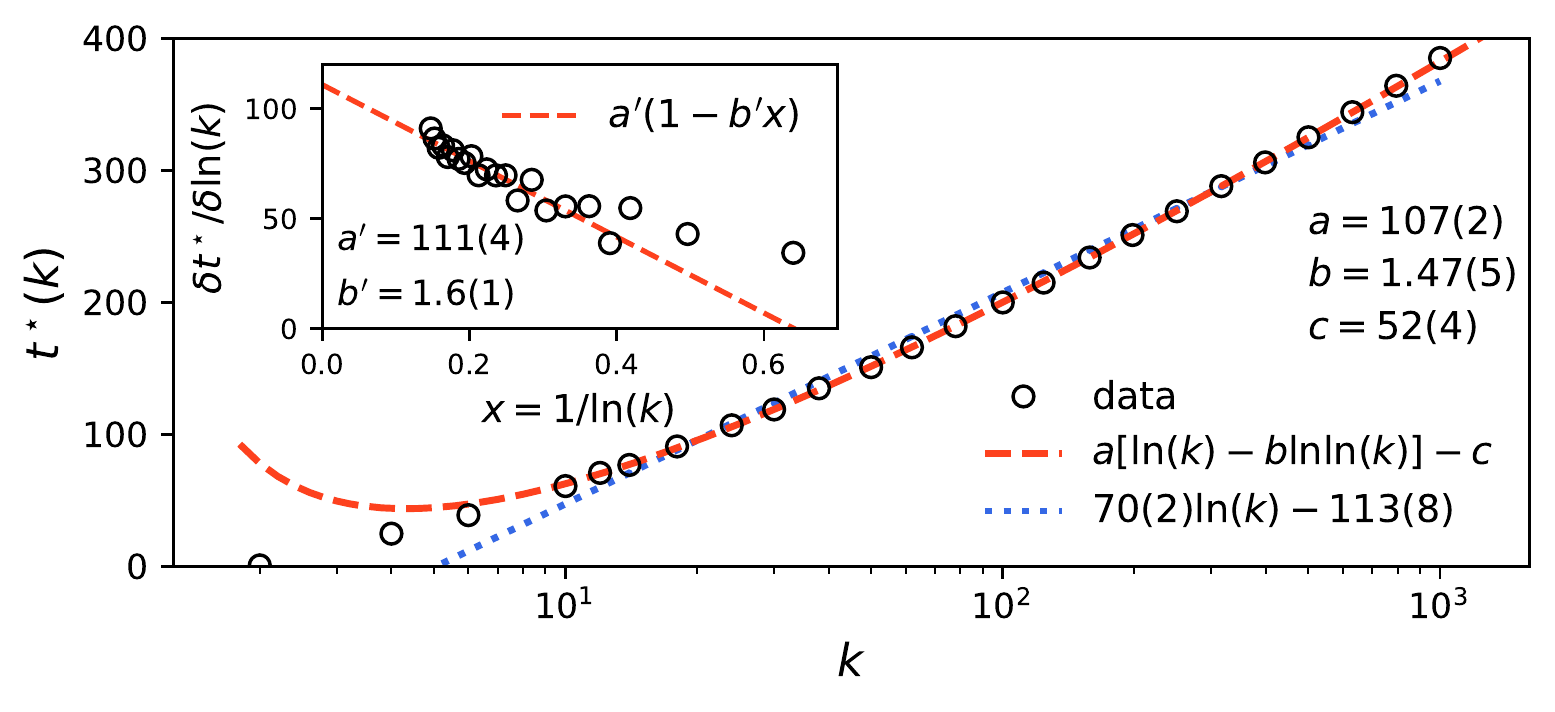}
    \caption{Optimal depth $t^\star(k)$ as a function of Pauli operator weight $k$, obtained from iMPS data as in Fig.~\ref{fig:itebd_shdn}, for $k$ up to 1000. The gate dilution is $\epsilon = 0.05$ and bond dimension is $\chi=2048$.
    Best fits to $t^\star(k)=a''\ln(k) - c''$ (dotted line) and $t^\star(k) = a[\ln(k)-b \ln\ln(k)] - c$ (dashed line) are shown. 
    The doubly-logarithmic correction is found to be $b = 1.47(5)$, consistent with the predicted $3/2$ in Eq.~\eqref{eq:opt_depth}. Inset: discrete derivatives $\delta t^\star(k) / \delta \ln(k)$, plotted vs $1/\ln(k)$, indicate a doubly-logarithmic correction $b = 1.6(1)$, also consistent with $3/2$.}
    \label{fig:tstar}
\end{figure}

{\it Numerical simulations.} 
To check the validity of the above results, we perform numerical simulations of the averaged twirling dynamics with infinite matrix product states (iMPS)~\cite{schollwock_density-matrix_2011} (see \cite{Note1}).
Fig.~\ref{fig:itebd_shdn}(a) shows the shadow norm for contiguous operators in a 1D chain of qubits ($q=2$), as a function of depth $t$.
Three regimes are clearly visible: 
the $t=0$ (local-twirling) value of $3^k$, 
a minimum at $t \sim \log(k)$,
and finally exponential growth due to continued operator spreading after relaxation. 
In un-diluted circuits ($\epsilon = 1$) the optimal depth $t^\star(k)$ takes very small integer values, severely limiting the resolution on its scaling~\cite{akhtar_scalable_2022}.
This issue is greatly alleviated by gate dilution: the shadow norm approximately behaves as a smooth function of an ``effective depth'' $\tau = \gamma(\epsilon)t$ (Fig.~\ref{fig:itebd_shdn}(b)), where $\gamma$ is the Pauli density relaxation rate in Eq.~\eqref{eq:n_bulk}---smaller $\epsilon$ yields a finer sampling of $\tau$.
To finely resolve the scaling of $t^\star(k)$, we set $\epsilon = 0.05$ obtaining the results in Fig.~\ref{fig:tstar}.
The data show remarkable agreement with Eq.~\eqref{eq:opt_depth}, including the subleading correction $\sim\ln \ln(k)$. 
The value of $3/2$ for the ratio of coefficients is universal (determined by the probability of first return of a random walk via Eq.~\eqref{eq:n_bulk}), which constitutes a nontrivial check of our analytical results.

{\it Higher dimensions.} 
While several details of the above discussion are special to one dimension, the general picture applies to systems in any finite spatial dimension.
The leading-order result ($t^\star(k) \sim \ln k$) depends only on the balancing of operator spreading and relaxation for operators whose boundary is much smaller than the bulk. 
In systems with all-to-all connectivity or on expander graphs, where a subsystem's bulk and boundary generally have comparable sizes, the optimal twirling depth is expected to be zero, i.e., local twirling performs best.
We test this expectation on a ``Brownian circuit'' model whose operator dynamics are described by simple, closed equations, and are amenable to exact treatment; we find the optimal depth is $t=0$ unless the operator is supported on a sufficiently large fraction of the system ($k \gtrsim N/2$), see~\cite{Note1}.


{\it Discussion.} 
We have studied how classical shadows based on shallow quantum circuits can be used to learn expectation values of Pauli operators. We have connected the sample complexity of classical shadows to the dynamics of operator weight, identifying two competing dynamical processes (operator spreading and relaxation) whose balance determines the optimal depth $t^\star$ of the twirling circuits. 
This picture elegantly explains previous numerical observations on one-dimensional systems~\cite{akhtar_scalable_2022, bertoni_shallow_2022}, and extends the result to systems in any finite dimension. Further, it shows that the optimal depth scales as $t^\star = O(\ln k)$ with the weight $k$ of the learned operator, as opposed to a more general $t^\star = {\rm polylog}(k)$ scaling~\cite{akhtar_scalable_2022}, ensuring a ${\rm poly}(k)$ classical computational cost for the optimal protocol.

Our work opens up several directions for future research. 
It would be interesting to generalize our results to different settings for classical shadows, beyond shallow brickwork circuits on qudits. The recent proposals for classical shadows in analog simulators~\cite{tran_measuring_2022, mcginley_shadow_2022} or on fermionic~\cite{zhao_fermionic_2021, wan_matchgate_2022} and bosonic~\cite{becker_classical_2022} systems are interesting possible directions. 
The validity of our results in higher dimension also suggests interesting applications to e.g. topological or fracton codes and phases~\cite{kitaev_fault-tolerant_2003, haah_local_2011, nandkishore_fractons_2019, pretko_fracton_2020, dua_sorting_2019, dua_clifford-deformed_2022}.
Further, it would be interesting to extend our analysis to measures of entanglement~\cite{brydges_probing_2019, elben_statistical_2019}, and to make contact with NISQ experiments~\cite{preskill_quantum_2018, struchalin_experimental_2021} by understanding the impact of noise on our results~\cite{chen_robust_2021, koh_classical_2022}.

Finally, the concept of operator relaxation may be of independent interest from the point of view of quantum dynamics.
While operator spreading is central to the study of quantum chaos~\cite{maldacena_bound_2016, swingle_measuring_2016, nahum_operator_2018, khemani_operator_2018, von_keyserlingk_operator_2018, rakovszky_diffusive_2018}, operator relaxation and similar diagnostics of local equilibration in operator space~\cite{rakovszky_diffusive_2018, khemani_operator_2018, omanakuttan_scrambling_2022} are comparatively under-explored, and may prove similarly useful in understanding signatures of quantum-chaotic behavior~\cite{gharibyan_onset_2018}.

\textit{Acknowledgments.} 
We thank Bryan Clark, Wen Wei Ho and Hsin-Yuan Huang for discussions on classical shadows.
M.I. and Y.L. are supported  in  part  by  the Gordon and Betty Moore Foundation's EPiQS Initiative through Grant GBMF8686. T.R and Y.L. are supported in part by the Stanford Q-FARM Bloch Postdoctoral Fellowship in Quantum Science and Engineering.  V.K. acknowledges support from the US Department of Energy, Office of Science, Basic Energy Sciences, under Early Career Award Nos. DE-SC0021111, the Alfred P. Sloan Foundation through a Sloan Research Fellowship, and the Packard Foundation through a Packard Fellowship in Science and Engineering. Numerical simulations were performed on Stanford Research Computing Center's Sherlock cluster. We acknowledge the hospitality of the Kavli Institute for Theoretical Physics at the University of California, Santa Barbara (supported by NSF Grant PHY-1748958).

\let\oldaddcontentsline\addcontentsline
\renewcommand{\addcontentsline}[3]{}
\bibliography{shadows}
\let\addcontentsline\oldaddcontentsline

\clearpage
\widetext

\setcounter{equation}{0}
\setcounter{figure}{0}
\setcounter{table}{0}
\setcounter{page}{1}
\makeatletter
\renewcommand{\thesection}{S\arabic{section}}
\renewcommand{\theequation}{S\arabic{equation}}
\renewcommand{\thefigure}{S\arabic{figure}}

\begin{center}
\textbf{\large Supplemental Material: Operator relaxation and the optimal depth of classical shadows} \\~\\
Matteo Ippoliti, Yaodong Li, Tibor Rakovszky, and Vedika Khemani \\
\textit{Department of Physics, Stanford University, Stanford, CA 94305}
\end{center}

\tableofcontents


\section{Derivation of shadow norm formula}

Here we provide techincal details involved in the derivation of our main result on the shadow norm of Pauli operators, Eq.~\eqref{eq:shdn_mainresult}. 

\subsection{Computation of the eigenvalues of $\mathcal{M}$}

Using the fact that Pauli operators are eigenmodes of $\mathcal{M}$~\cite{akhtar_scalable_2022, bertoni_shallow_2022}, $\mathcal{M}(O_A) = \lambda_A O_A$, and the expression for the measurement channel with locally-scrambled twirling ensemble
\begin{equation}
    \mathcal{M}[O_A] = D \int{\rm d}U\ U^\dagger \ketbra{0}{0} U\ \bra{0}U O_A U^\dagger \ket{0},
\end{equation}
(where we used the locally-scrambled property to replace each bitstring state $\ket{b}$ with $\ket{0}$, hence the factor of Hilbert space dimension $D$).
we can compute the eigenvalues $\lambda_A$ via
\begin{equation}
    \lambda_A 
    = \frac{1}{D} {\rm Tr}(O_A^\dagger \mathcal{M}[O_A])
    = \int {\rm d}U \left| \bra{0} U O_A U^\dagger \ket{0}\right|^2 .
    \label{eq:supp_lambdaR}
\end{equation}
Next, we write $U O_A U^\dagger \equiv O_A(t)$, and expand it in the Pauli basis: $O_A(t) = \sum_P \alpha_P(t) P$, where $P$ ranges over the whole $N$-qudit Pauli group and $\alpha_P(t) = {\rm Tr}(P^\dagger O_A(t))/D$. 
Furthermore, we exploit the locally-scrambled property of the twirling ensemble to replace $\ket{0}$ with a Haar-random product state, $\ket{\Psi} = \bigotimes_{i=1}^N \ket{\psi_i}$, with each $\ket{\psi_i}$ an independent Haar-random state of a $q$-state qudit. 
We obtain
\begin{equation}
    \lambda_A 
    =  \int{\rm d}U\ {\rm d}\Psi \left| \bra{\Psi} \left( \sum_P \alpha_P(t) P \right) \ket{\Psi}\right|^2
    = \sum_{P,P'} \int{\rm d}U \alpha_P^\ast(t) \alpha_{P'}(t)\int {\rm d}\Psi  \bra{\Psi} P^\dagger  \ket{\Psi} \bra{\Psi} P' \ket{\Psi} \label{eq:supp_lambda_integral}
\end{equation}
Owing to random dephasing, terms in the sum with $P \neq P'$ vanish.
To see this, let us assume that $P$ and $P'$ differ at some site $i$, where $P|_i = X^{m_1} Z^{n_1}$ and $P'|_i = X^{m_2} Z^{m_2}$ (recall $Z$ and $X$ are the `clock' and `shift' operators which generate the generalized Pauli group) with $(m_1,n_1) \neq (m_2, n_2)$. If $m_1\neq m_2$, we can redefine $\ket{\Psi} \mapsto Z_i \ket{\Psi}$ (since the measure ${\rm d}\Psi$ is invariant under local unitaries). Using the algebra of clock and shift operators, $X_i Z_i = Z_i X_i e^{2\pi i/q}$, we see that this transforms the integrand as 
\begin{equation}
\bra{\Psi} P^\dagger \ketbra{\Psi}{\Psi} P' \ket{\Psi} 
\mapsto \bra{\Psi} Z_i^\dagger P^\dagger Z_i \ketbra{\Psi}{\Psi} Z_i^\dagger P' Z_i \ket{\Psi} 
= \bra{\Psi} P^\dagger \ketbra{\Psi}{\Psi} P' \ket{\Psi} e^{2\pi i (m_2 - m_1) / q}.
\end{equation}
Thus the integral vanishes unless $m_1 = m_2$. The same argument (with $\ket{\Psi} \mapsto X_i \ket{\Psi}$) forces $n_1 = n_2$. 
Thus
\begin{equation}
    \lambda_A = \sum_P \overline{|\alpha_P(t)|^2} \int{\rm d}\Psi\ |\bra{\Psi}P\ket{\Psi}|^2
    = \sum_P \overline{|\alpha_P(t)|^2} (q+1)^{-|P|},
\end{equation}
where $|P|$ denotes the Hamming weight of $P$, i.e. the number of non-identity Pauli matrices in $P$, and $\overline{[\cdots]} \equiv \int {\rm d}U\ [\cdots] $.
The result follows from the single-qudit Haar integral 
\begin{equation} 
\int {\rm d}\psi_i |\bra{\psi_i} O_i \ket{\psi_i}|^2 = 
    \left\{ \begin{aligned}
        & 1/(q+1) \text{ if } O_i \text{ is a traceless Pauli matrix,} \\
        & 1 \text{ if } O_i \text{ is the identity.}
    \end{aligned} \right.
\end{equation}
Finally, by collecting all Pauli strings of the same Hamming weight in the sum, we arrive at the result:
\begin{equation}
    \lambda_A = \sum_w (q+1)^{-w} \sum_{P: |P| = w} \overline{|\alpha_P(t)|^2}
    = \sum_w (q+1)^{-w} \pi_{A,t}(w),
\end{equation}
where the last equality defines the weigth distribution $\pi_{A,t}(w)$.
We note that Ref.~\cite{qi_measuring_2019} derives the same formula from computing the variance of an expectation value $\bra{\Psi} O \ket{\Psi}$ over the ensemble of random product states $\ket{\Psi} = \bigotimes_i \ket{\psi_i}$ for a fixed operator $O(t)$. In our case the operator $O_A(t)$ is not fixed but is itself random (due to the twirling), however one may apply the derivation in Ref.~\cite{qi_measuring_2019} to the integrand in Eq.~\eqref{eq:supp_lambda_integral} and subsequently average over ${\rm d}U$. 

\subsection{Shadow norm for Pauli operators}

Here we derive the identity $\shadownorm{O_A}^2 = \lambda_A^{-1}$ for Pauli operators. Note that this was already observed in Ref.~\cite{bertoni_shallow_2022} and, for the state-averaged shadow norm, in Ref.~\cite{akhtar_scalable_2022}. We include a derivation here for the sake of clarity and completeness.

We begin by showing that $\shadownorm{O_A}^2$ is independent of the underlying state $\rho$ under the assumption of local scrambling, when $O_A$ is a Pauli operator (see also Ref.~\cite{bertoni_shallow_2022}). 
The variance of the estimator $\hat{o}$ is comprised of two terms: $\mathbb{E}[|\hat{o}|^2]$ and $|\mathbb{E}[\hat{o}]|^2$ (note $\hat{o}$ may be complex for $q>2$, as generalized Pauli operators are unitary but generally not Hermitian). 
The latter term is equal to $|\langle O_A \rangle|^2$ and thus of order 1 for a Pauli operator. We focus on the former:
\begin{equation}
    \mathbb{E}_{U,b}[|\hat{o}|^2]
    = \int {\rm d}U \sum_b \bra{b}U \rho U^\dagger \ket{b} \left| {\rm Tr}[O_A \mathcal{M}^{-1}(U^\dagger \ketbra{b}{b}U)] \right|^2.
\end{equation}
By using the fact that $\mathcal{M}^{-1}$ is self-adjoint (evident from the fact that it has an orthonormal eigenbasis with real eigenvalues), and that $\mathcal{M}^{-1}(O_A) = \lambda_A^{-1} O_A$, we arrive at 
\begin{equation}
    \mathbb{E}_{U,b}[ |\hat{o}|^2]
    = \lambda_A^{-2} D \int{\rm d}U \bra{0}U\rho U^\dagger \ket{0} |\bra{0}UO_AU^\dagger \ket{0}|^2,
\end{equation}
where we again used that the ensemble of unitaries is locally scrambled. 
We now expand $\rho = \sum_P c_P P$, where $P$ runs over all $q^N$ Pauli operators. This gives
\begin{equation}
    \mathbb{E}_{U,b} [|\hat{o}|^2]
    = \sum_P c_P \lambda_A^{-2} \int{\rm d}U \bra{0} UPU^\dagger\ket{0} |\bra{0}UO_AU^\dagger \ket{0}|^2.
\end{equation}
If $P\neq I$, the integral vanishes due to random dephasing: there exists a single-site (generalized) Pauli unitary $u_i$ such that $u_i P u_i^\dagger = e^{i\phi}P$ with a nonzero phase $\phi$, while the measure ${\rm d}U$ is invarant under $U\mapsto U u_i$ and any phase accumulated by $O_A$ cancels out due to the absolute value; thus the integral must vanish. 

For this reason, the only term in $\rho$ which contributes to the shadow norm is $I/D$, which gives
\begin{equation}
    \mathbb{E}_\rho \mathbb{E}_{U,b} (\hat{o}^2)
    = \lambda_A^{-2} \int{\rm d}U |\bra{0}UO_AU^\dagger \ket{0}|^2
    = \lambda_A^{-1}
\end{equation}
where the last equality follows from recognizing the expression in Eq.~\eqref{eq:supp_lambdaR} for $\lambda_A$.

\subsection{Local and global twirling}

Here we show how to recover the well-known results for local and global twirling~\cite{huang_predicting_2020} from Eq.~\eqref{eq:shdn_mainresult}.
The local twirling case is recovered for $t = 0$ (note $t$ measures the number of two-qudit gates in the circuit, but initial and final layers of random one-qudit Clifford gates is always assumed). We have $\pi_{A,0}(w) = \delta_{w,k}$ (single-qudit gates cannot change the weight of a Pauli operator), thus 
\begin{equation}
\shadownorm{O_A}^2 = \left[ \sum_w \pi_{A,0}(w) (q+1)^{-w}\right]^{-1} = (q+1)^k \;,
\end{equation}
recovering the well-known $3^k$ result for qubits ($q=2$).
Global twirling is recovered for $t\to\infty$.
In this limit $O_A(t)$ becomes a random (traceless) Pauli operator. Then, the weight distribution approximately factors across different sites, with each site hosting a random Pauli matrix (the only correlation comes from removing the global identity operator; this can be neglected for large $N$). 
As there are $q^2$ Pauli matrices (including the identity), we have an ``empty'' site (i.e. an identity) with probability $q^{-2}$ and an ``occupied'' site (i.e. a traceless Pauli) otherwise. 
Thus we have
\begin{equation}
    \| O \|_\text{sh,avg}^{-2} 
    = \prod_i \sum_{n_i=0,1} (q+1)^{-n_i} {\sf Prob}(n_i)
    = \prod_i \left( q^{-2} + \frac{1-q^{-2}}{q+1} \right)
    = q^{-N} = \| O \|_F^{-2},
\end{equation}
i.e. the shadown norm equals the Frobenius norm, as expected for global twirling.


\section{Random walk picture for operator relaxation \label{sec:supp_rw}}

Here we present a detailed discussion of the mapping of operator relaxation to a random-walk problem. 

\subsection{Update rules and mapping to random walk}

We aim to describe the evolution of the vector $p_{\mathbf n}$ of probabilities of occupation configurations $\mathbf n \in \{0,1\}^N$, where $n_i$ represents the occupation of site $i$ in the (identity, traceless Pauli) basis. 

The identity on two qudits is invariant under any unitary, thus $|\vacuum\vacuum) \mapsto |\vacuum\vacuum)$; 
under Haar-random two-qudit gates, any traceless Pauli operator maps onto one of the $q^4-1$ traceless Pauli operators with equal probability, thus $|\vacuum\occupied) \mapsto a|\vacuum\occupied) + a |\occupied\vacuum) + (1-2a) |\occupied\occupied)$, etc, where $a = 1/(q^2+1)$ (the fraction of single-qudit traceless Paulis out of all two-qudit traceless Paulis, $(q^2-1)/(q^4-1)$). 
It follows that the two-site update matrix in the $\{ |\vacuum\vacuum), |\vacuum\occupied), |\occupied\vacuum), |\occupied\occupied)\}$ basis reads \begin{equation}
\label{eq:supp_M_haar} 
    M_\text{Haar}
    = \begin{pmatrix} 
        1 & 0 & 0 & 0 \\
        0 & a & a & a \\
        0 & a & a & a \\
        0 & 1-2a & 1-2a & 1-2a
        \end{pmatrix}    
\end{equation}
Introducing the vector $|\free) = |\vacuum) + |\occupied)$, from the fact that $M_\text{Haar}$ is a stochastic matrix (i.e. columns add up to unity) we immediately have $M_\text{Haar}^T |\free\free) = |\free\free)$, thus a domain of $\free$ states is an eigenstate of the backwards dynamics.
Furthermore, we have by explicit calculation that $M_\text{Haar}^T |\free\vacuum) = (1-a)|\vacuum\vacuum) + a|\free\free)$.
This implies that, under the backward evolution, domains of $\vacuum$ and $\free$ states are preserved, and the location of the domain wall hops by one site either left or right, with probabilities $a$, $1-a$ that favor the growth of the $\vacuum$ domain (as $a=1/(q^2+1)<1/2$).

In the presence of gate dilution, we have $M(\epsilon) = (1-\epsilon)I + \epsilon M_\text{Haar}$, and thus the update rule for a domain wall between $\free$ and $\vacuum$:
\begin{equation}
    |\overbrace{\free\cdots\free}^x \overbrace{\vacuum\cdots\vacuum}^{N-x})
    \mapsto 
    \epsilon a |\overbrace{\free\cdots\free}^{x+1} \overbrace{\vacuum\cdots\vacuum}^{N-x-1}) + 
    (1-\epsilon) |\overbrace{\free\cdots\free}^{x} \overbrace{\vacuum\cdots\vacuum}^{N-x}) + 
    \epsilon (1-a) |\overbrace{\free\cdots\free}^{x-1} \overbrace{\vacuum\cdots\vacuum}^{N-x+1}).
    \label{eq:supp_rw_hop}
\end{equation}
This update is illustrated in Fig.~\ref{fig:rw}(a) in the main text.

\subsection{Survival probability}

For the purpose of computing the average density of holes $\overline{h_i}(t)$, the final boundary condition (Fig.2(a) in the main text) is $|\cdots \free\free\free \vacuum \free \free\free \cdots)$, where $\vacuum$ is at site $i$ and the 1D chain is infinite (though in practice length $t$ on both sides suffices, due to the unitary light cone). 
This corresponds to two random walkers being initialized at positions $x_1 = i$, $x_2 = i+1$ (bond $i$ being to the left of site $i$). 

Let us focus on $\epsilon = 0$ (un-diluted circuit). The relative coordinate $x_r \equiv x_2 - x_1$ takes exactly two steps per unit time (as each of the two walkers $x_{1,2}$ takes one step), with ${\sf Prob}(\delta x=+1) = 1-a$ and ${\sf Prob}(\delta x = -1) = a$. The walker is annihilated if it reaches $x_r = 0$ in $t$ steps or less, it survives otherwise. 
If it survives, then the contribution to $\overline{h_i}(t)$ includes overlaps $(\vacuum|\occupied) = 0$, and thus vanishes (see Fig.~\ref{fig:rw}(b) in the main text). 
Thus $\overline{h_i}(t)$ equals the probability of annihilation at or before time $t$. 

The probability of annihilation at time $\tau$, i.e. in $2\tau + 1$ steps (note that at $t=0$ only one out of $x_{1,2}$ takes a step) is given by
\begin{equation}
\label{eq:P-ann-at-time-tau}
    P(\text{annihilation at time } \tau) = a^{\tau+1} (1-a)^\tau C_\tau \simeq \frac{a[4a(1-a)]^\tau }{\sqrt{\pi \tau^3}}
\end{equation}
where $C_\tau = \frac{1}{\tau+1}\binom{2\tau}{\tau}$ are the Catalan numbers, whose large-$\tau$ expansion is $C_\tau \simeq 4^\tau / \sqrt{\pi \tau^3}$. 
Defining 
\begin{equation} 
e^{-\gamma} \equiv 4a(1-a) = \left( \frac{2q}{q^2+1} \right)^2  \label{eq:supp_gamma}
\end{equation} 
(note $\gamma>0$) and integrating over $\tau\leq t$, we see that at large $t$,
\begin{equation}
    P(\text{annihilation at time}\leq t) = {\rm const.} - \frac{a}{\gamma\sqrt{\pi}} t^{-3/2} e^{-\gamma t}
\end{equation}
up to subleading corrections. 
We can determine the integration constant as follows:
by reversing the bias ($a\mapsto 1-a$), the walker should annihilate with unit probability as $t\to\infty$; this implies $\sum_\tau a^\tau (1-a)^{\tau + 1} C_\tau = 1$; exploiting this identity, we have that the late-time asymptotic value of the annihilation probability is $\sum_\tau a^{\tau+1} (1-a)^\tau C_\tau = \frac{a}{1-a} = \frac{1}{q^2}$. This is in agreement with the expected density of identity operators $\overline{h_i}$ at late times.
With this, we conclude that 
\begin{equation}
    \overline{n_i}(t) = P(\text{survival up to } t) \simeq 1 - \frac{1}{q^2} + \frac{1}{\sqrt{\pi} \gamma (q^2+1)} t^{-3/2} e^{-\gamma t}
\end{equation}
at large $t$, up to subleading corrections.

The functional form $t^{-3/2} e^{-\gamma t}$ is a universal property of a biased random walk's first return time, and thus holds with gate dilution as well, though the value of $\gamma$ and the multiplicative constant will change.

\subsection{Higher dimension \label{sec:supp_2d}}

\begin{figure}[h]
    \centering
    \includegraphics[width=.4\textwidth]{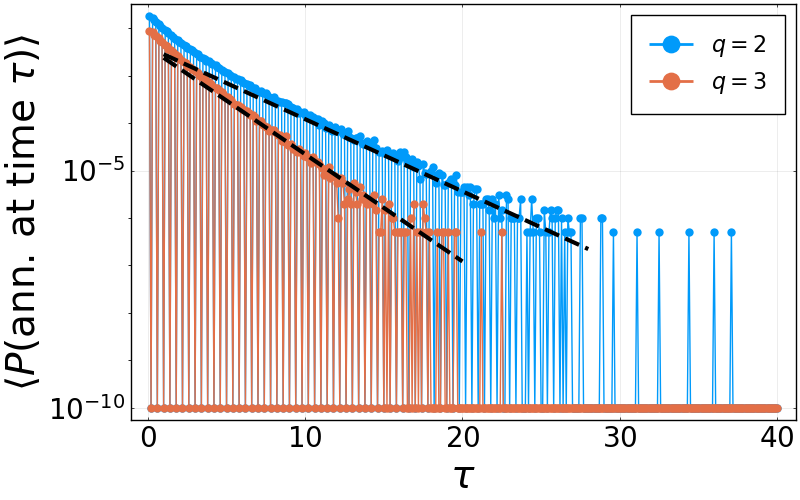}
    \caption{Numerical results for the probability of annihilation of the $\vacuum$ domain at time $\tau$, for qudits with $q = 2$ and 3 in two spatial dimensions. Data from Monte Carlo simulation of the Markov process in the $\vacuum$, $\free$ basis, without gate dilution ($\epsilon = 1$). Dashed lines are exponential fits.}
    \label{fig:2d-RW}
\end{figure}

Although the random walk picture only strictly works in one dimension, the Markov process in Eq.~\eqref{eq:supp_M_haar} is defined for any circuit with 2-local gates, and in particular can be used to compute the mean operator density in higher dimensions.
Here we simulate this process in two dimensions, with the state $|\vacuum)$ on a single site, and the state $|\free)$ on all remaining sites.
The boundary between the two domains, when present, moves under Eq.~\eqref{eq:supp_rw_hop}.
We plot $P(\text{annihilation at time}\  \tau)$ in Fig.~\ref{fig:2d-RW}.
The result is similar to the one dimensional case, Eq.~\eqref{eq:P-ann-at-time-tau}, namely to leading order
\begin{align}
    P(\text{annihilation at time}\  \tau) \propto e^{-\gamma(q) \tau}.
\end{align}
We expect this result to hold in general in any finite dimension.
As a consequence, the leading-order result $t^\star \propto \ln k$ should hold in all finite dimensions, with possibly different subleading corrections.


\section{Minimization of the upper bound \label{sec:supp_ub}}

Here we show details on the minimization of the upper bound to the shadow norm, $\shadownorm{O_A}^2 \leq (q+1)^{\overline{w}}$.
We aim to find the minimum over $t$ of the average weight $\overline{w}(t)$, which controls the upper bound $\shadownorm{O_A}^2 \leq (q+1)^{\overline{w}}$.
The average spatial length of the twirled Pauli operator $O_A(t)$ is $\overline{\ell} (t) = k + 2v_B t$, where $v_B$ is the average butterfly velocity of the unitary circuit $U$.
Neglecting the front structure of the operator (i.e. dependence of the density $\overline{n_i}$ on position $i$ inside the support of the operator), we have $\overline{w} \simeq \overline{\ell}(t) \overline{n_i}(t)$, where $\overline{n_i}(t) = 1-1/q^2+ c f(t)$ for $c>0$ constant and $f(t) = t^{-3/2} e^{-\gamma t}$.
Minimization of $\overline{w}$ in this approximation gives
\begin{equation}
    2v_B \left[1-1/q^2+ c f(t) \right] = - (k + 2v_B t) cf'(t).
\end{equation}
We are interested in the asymptotics for large Pauli operators and late times ($k,t\gg 1$),
thus we may drop $cf(t)$ in the l.h.s. (as it is $\ll 1$), and in the r.h.s. we may simplify $f'(t) = -[\gamma + \frac{3}{2t}] f(t) \simeq -\gamma f(t)$; thus
\begin{equation}
    2v_B (1-1/q^2) \simeq \gamma c (k+2v_Bt) t^{-3/2} e^{-\gamma t}.
\end{equation}
It follows (since the l.h.s. is finite) that the minimum is achieved for $t \sim \log(k)$; thus $k\gg t \gg 1$. Dropping $2v_Bt$ in the r.h.s., we get
\begin{equation}
    t^{3/2} e^{\gamma t} \simeq \frac{\gamma}{2v_B(1-1/q^2)} k.
\end{equation}
Taking the logarithm, we arrive at the recursive equation 
\begin{equation}
    t^\star = \frac{1}{\gamma} \left( \ln(k) - \frac{3}{2} \ln(t^\star) \right) + {\rm const.} \label{eq:supp_tstar}
\end{equation}
Iterating the recursion once (i.e. setting $t^\star\mapsto \frac{1}{\gamma}\ln(k) + \dots$ in the r.h.s.) yields the result in the main text, Eq.~\eqref{eq:opt_depth}.

At depth $t^\star(k)$, the upper bound reads
\begin{equation}
    \shadownorm{O_A}^2 \leq (q+1)^{(k+O(\log(k)))(1-q^{-2} + O(1/k))} = (q+1)^{(1-q^{-2})k} \times {\rm poly}(k).
\end{equation}


\section{Mean-field approximation to the shadow norm \label{sec:supp_meanfield}}

Here we present details about the analytical mean-field approximation to the shadow norm.
The key idea is to drop correlations between Pauli densities at different sites, taking the $n_i$ to be independently, identically distributed (i.i.d.) binomial random variables, in order to factor the average $\overline{(q+1)^{-w}}$ into a product of on-site averages.
However, there is an important subtlety in how to treat the size of the operator's support, as explained below.

\subsection{Saddle-point bulk density \label{sec:supp_nsp}}

While we have computed the {\it average} bulk density of Paulis inside a long operator $\overline{n_i}(t)$, Eq.~\eqref{eq:n_bulk}, the dominant contribution to the shadow norm comes from Pauli strings at below-equilibrium density.
This is due to the exponential suppression induced by the $(q+1)^{-w}$ factor.
Let us consider a Pauli operator on $N$ sites, with $\overline{n_i}(t) \equiv 1-q^{-2} + cf(t)$ for all sites $i$; 
within a mean-field approximation, treating each $n_i$ as an i.i.d. binomial distribution with ${\sf Prob}(n_i = +1) = \overline{n_i}(t)$, we have
\begin{equation}
    \overline{(q+1)^{-w}} 
    = \sum_{w=1}^N \binom{N}{w} (1-q^{-2}+cf(t))^w (q^{-2}-cf(t))^{N-w} (q+1)^{-w}
    = \left( \frac{1}{q} -\frac{q}{q+1} cf(t) \right)^N.
\end{equation}
We define a ``saddle point'' density $n_{\rm sp}$ by setting the above equal to $(q+1)^{-N n_{\rm sp}}$ (in other words, $n_{\rm sp}$ is defined so that if the weight distribution were a $\delta$-function centered at $w = Nn_{\rm sp}$, we would recover the correct answer for $\overline{(q+1)^{-w}}$).
This gives
\begin{equation}
    n_{\rm sp}  = 
    \frac{\ln(q)}{\ln(q+1)} 
    - \frac{\ln(1-q^2cf(t)/(q+1))}{\ln(q+1)} 
    \simeq \frac{\ln(q)}{\ln(q+1)} 
    + \frac{q^2cf(t) }{(q+1) \ln(q+1)}
\end{equation}
where we linearized in $f(t) \ll 1$ in the second step. 
Note that $n_{\rm sp} < \overline{n}$; in particular for large $q$ we have $n_{\rm sp}\simeq 1-1/q$ vs $\overline{n} = 1-1/q^2$.

\subsection{Mean-field result \label{sec:supp_mfr}}

Let us now consider the biased random walk of an operator's front, $x$. 
This moves outward with probability $1-a$ and inward otherwise, giving a butterfly velocity $v_B = 1-2a$ for the {\it average} position of the front. 
Let us consider the left and right endpoints of the operator, $x_1$ and $x_2$, each describing a random walk as above (with average velocity $\pm v_B$).
Assuming that all sites between the two fronts host an i.i.d. local density $n_i$ with mean $\overline{n_i}$, we find that the contribution of such an operator to the (inverse squared) shadow norm $\shadownorm{O_A}^{-2}$ is $(q+1)^{n_{\rm sp}(x_1-x_2)}$, by the definition of $n_{\rm sp}$ given above.
As long as $k \gg t$, the two endpoints cannot meet and we may treat the random walks as independent; thus
\begin{equation}
    \shadownorm{O_A}^{-2} 
    \simeq \sum_{x_1, x_2} {\sf Prob}(x_1) {\sf Prob}(x_2) (q+1)^{-n_{\rm sp}(x_2-x_1)}
    = (q+1)^{-n_{\rm sp}k} \left( \sum_{x=-t}^t {\sf Prob}(x) (q+1)^{-n_{\rm sp}x} \right)^2 
\end{equation}
where $x$ describes the outward displacement of either endpoint from its initial location.
The sum in parenthesis yields
\begin{equation}
\sum_{x=-t}^{+t} \binom{t}{\frac{t+x}{2}} a^{\frac{t-x}{2}} (1-a)^{\frac{t+x}{2}} (q+1)^{-n_{\rm sp} x}
= \left( a (q+1)^{n_{\rm sp}} + (1-a) (q+1)^{-n_{\rm sp}} \right)^t \label{eq:supp_shdn_vbsp}
\end{equation}
Setting this equal to $q^{-v_B^{\rm sp}t}$ defines the ``saddle point butterfly velocity''
\begin{equation}
v_B^{\rm sp} = \frac{\ln(ag + (1-a)/g)}{\ln q},
\qquad
g = (q+1)^{n_{\rm sp}} = q \left(1-\frac{q^2cf(t)}{q+1}\right)^{-1}
\end{equation}
We note that at equilibrium, i.e. for $t\to\infty$ ($f(t) \to 0$), this gives 
\begin{equation} 
v^{\rm sp}_B = \log_q \frac{q^2+1}{2q}, 
\label{eq:supp_vbsp}
\end{equation}
equal to the entanglement velocity $v_E$~\cite{nahum_quantum_2017} and proportional to the Pauli density relaxation rate $\gamma = 2\ln \frac{q^2+1}{2q}$, Eq.~\eqref{eq:supp_gamma}; namely we have $q^{2v_B^{\rm sp}} = e^\gamma$.

\begin{figure}
    \centering
    \includegraphics[width=\textwidth]{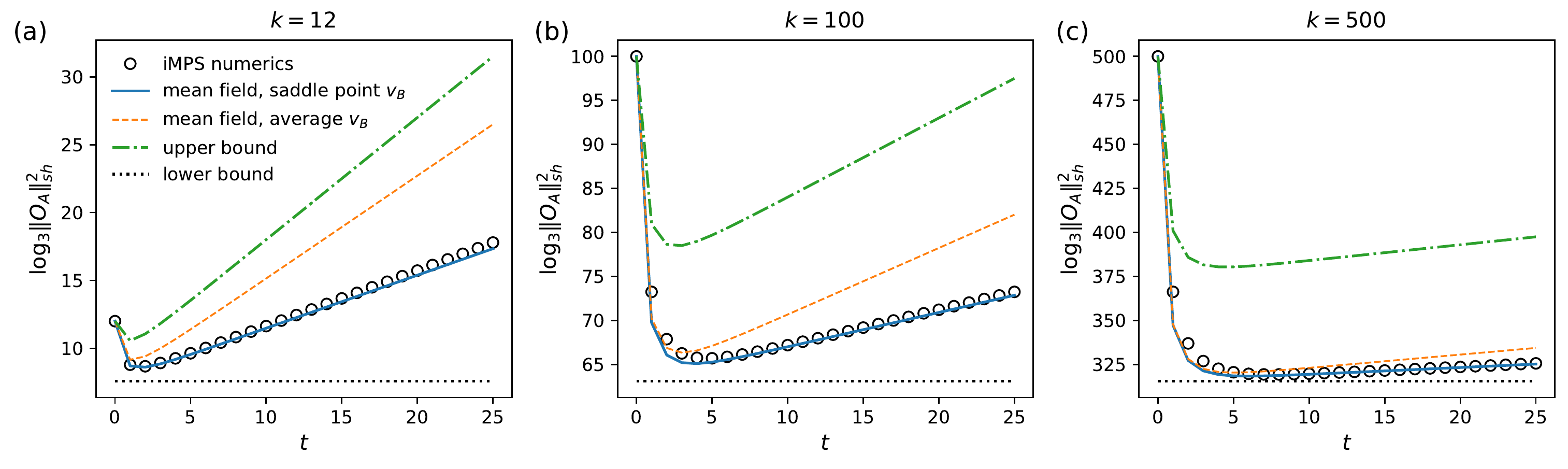}
    \caption{Comparing the mean-field result for $\shadownorm{O_A}^2$, Eq.~\eqref{eq:supp_meanfield_shdn} (solid line), to iMPS numerical data (circles), the upper bound $\shadownorm{O_A}^2 \leq (q+1)^{\overline{w}}$ (dot-dashed line), the lower bound $\shadownorm{O_A}^2 \geq q^k$ (dotted line), and an alternative mean-field approximation where the average butterfly velocity $v_B$ is used instead of the saddle-point value $v_B^{\rm sp}$ (dashed line).}
    \label{fig:supp_meanfield}
\end{figure}

With this notation, we may rewrite Eq.~\eqref{eq:supp_shdn_vbsp} to obtain an explicit mean-field approximation to the shadow norm,
\begin{equation}
    \shadownorm{O_A}^{2} \simeq (q+1)^{n_{\rm sp}k} q^{2v_B^{\rm sp}t}
    = \left( \frac{1}{q} - \frac{qcf(t)}{q+1} \right)^{-k} q^{2v_B^{\rm sp} t}.
    \label{eq:supp_meanfield_shdn}
\end{equation}
Notably, at late times this predicts
\begin{equation}
\shadownorm{O_A}^2 \simeq q^{k + 2v_B^{\rm sp} t} \simeq q^k e^{\gamma t}, 
\end{equation}
where we used the fact that $v_B^{\rm sp} \to \gamma/ 2\ln(q) $ at late times, Eq.~\eqref{eq:supp_vbsp}.
This is found to be in quantitative agreement with numerical data, see Fig.~\ref{fig:supp_meanfield}.

\subsection{Optimal depth \label{sec:supp_optdepth}}

Finally we minimize Eq.~\eqref{eq:supp_meanfield_shdn} to obtain the mean-field prediction for the optimal depth:
\begin{equation}
    [2v_B^{\rm sp} + 2t\partial_t v_B^{\rm sp}]  \ln \left(\frac{1}{q} - \frac{qcf(t)}{q+1} \right) - (k+2v_B^{\rm sp}t) \frac{qcf'(t)}{1+q^{-1} - qcf(t)} = 0
\end{equation}
Discarding small terms for $k\gg t\gg 1$ we arrive at 
\begin{equation}
    kf(t) = \frac{2v_B^{\rm sp} (1+1/q) \ln(q)}{\gamma q c}
\end{equation}
which upon taking the logarithm of both sides yields Eq.~\eqref{eq:supp_tstar}, and thus the same optimal depth $t^\star(k) = \frac{1}{\gamma} (\ln(k) - (3/2)\ln\ln(k) + \dots)$.
The mean-field shadow norm at $t^\star(k)$ is $q^k\times {\rm poly}(k)$.


\section{Computation of velocity scales with gate dilution \label{sec:supp_velocities}}

Here we show numerical and analytical results for the various velocity scales in the problem in the presence of gate dilution, $0\leq \epsilon<1$.  
The relevant scales are: 
$\gamma$, ``relaxation rate'' of the Pauli density; 
$v_B$, butterfly velocity, governing operator spreading on average;
$v_B^{\rm sp}$, governing operator spreading at the saddle point in the shadow norm calculation;
and $v_E$, entanglement velocity, characterizing entanglement growth.
Their values in un-diluted Haar-random brickwork circuits ($\epsilon = 1$) are easily computed analytically. 
We have 
$v_B = \frac{q^2-1}{q^2+1}$, 
$v_E = v_B^{\rm sp} = \log_q \frac{q^2+1}{2q}$ (see Eq.~\eqref{eq:supp_vbsp}), and
$\gamma = 2\ln\frac{q^2+1}{2q} = 2\ln(q) v_E$.

\begin{figure}
    \centering
    \includegraphics[width=\textwidth]{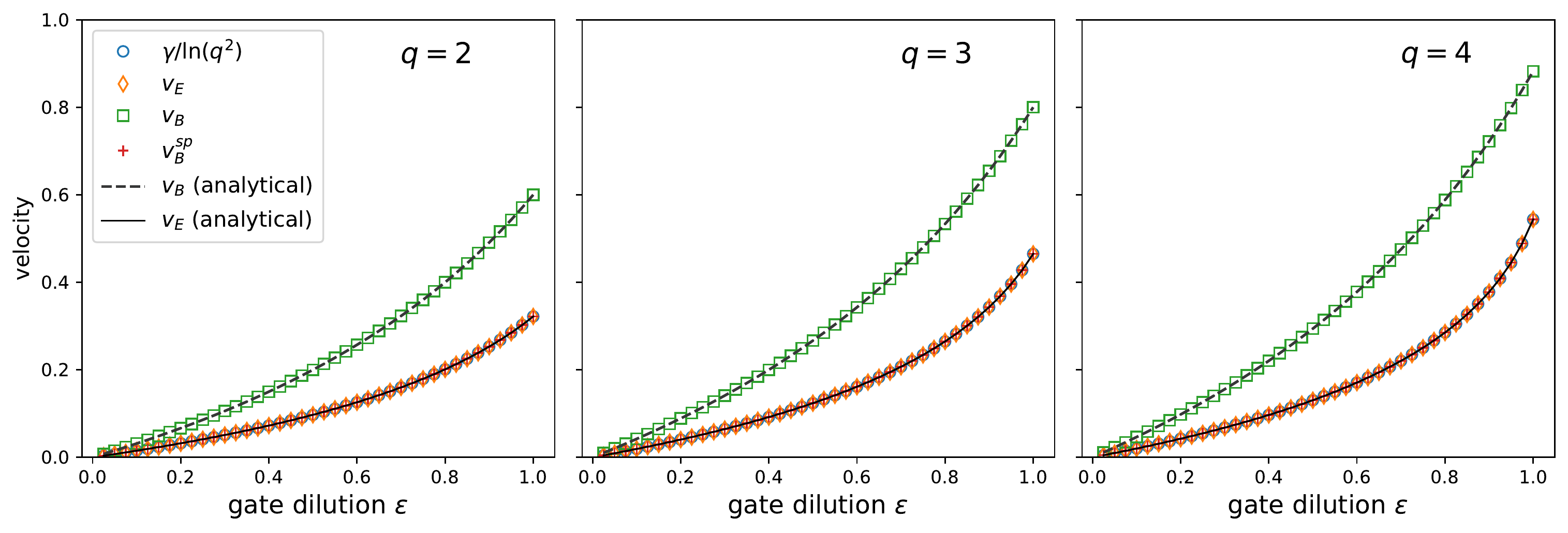}
    \caption{Velocity scales $\gamma$ (Pauli density relaxation rate), $v_E$ (entanglement velocity), $v_B$ (butterfly velocity), and $v_B^{\rm sp}$ (saddle-point butterfly velocity) computed numerically as a function of gate dilution $\epsilon$, for qudit dimension $q=2$ (left), 3 (center), 4 (right). 
    $v_B$ and $v_E$ agree with their respective analytical predictions, Eq.~\eqref{eq:supp_vb_eps} and Eq.~\eqref{eq:supp_ve_eps}. Furthermore, we find that within numerical error, $v_E = v_B^{\rm sp} = \gamma / \ln(q^2)$ across all values of $\epsilon$ and $q$, as predicted. }
    \label{fig:velocities}
\end{figure}

The corresponding values for $0 < \epsilon \leq 1$ and several local Hilbert space dimensions $q$ are shown in Fig.~\ref{fig:velocities}. The methods used for the computation are described below for each quantity.
Remarkably, we find that the equality $v_E = v_B^{\rm sp} = \gamma / \ln(q^2)$, initially derived at $\epsilon = 1$, persists to $\epsilon < 1$, suggesting that this might hold more generally (i.e. beyond diluted Haar-random brickwork circuits).

\vspace{0.5cm}
{\bf Butterfly velocity.}
$v_B$ is the average velocity of the random walk described by the right endpoint of a Pauli operator, i.e. the rightmost site $x$ hosting a traceless Pauli matrix (of course the left endpoint has average velocity $-v_B$). 
The transition probabilities for $x$ depend on the parity of $x+t$, due to the brickwork structure of the circuit. We have
\begin{equation}
\text{if } x+t \text{ is even:}\
\left\{ 
\begin{aligned}
& {\sf Prob}(\delta x=+1) = \epsilon(1-a)\\
& {\sf Prob}(\delta x=0) = 1-\epsilon + \epsilon a
\end{aligned} \right. ,
\qquad
\text{if } x+t \text{ is odd:}\
\left\{ 
\begin{aligned}
& {\sf Prob}(\delta x=0) = 1-\epsilon a\\
& {\sf Prob}(\delta x=-1) = \epsilon a
\end{aligned} \right. \label{eq:supp_vb_rw}
\end{equation}
From these transition rules we find that the steady-state probability of $x+t$ being even is ${\sf Prob}(x+t\text{ even}) = (1-\epsilon a) / (2-\epsilon)$, and thus
\begin{align}
v_B(\epsilon) 
& = \overline{\delta x} = {\sf Prob}(x+t\text{ even}) {\sf Prob}(\delta x=+1|x+t\text{ even})
- {\sf Prob}(x+t\text{ odd}) {\sf Prob}(\delta x=-1|x+t\text{ odd}) \nonumber \\ 
& = \frac{\epsilon}{2-\epsilon}(1-2a) = \frac{\epsilon}{2-\epsilon} v_B(1).
\label{eq:supp_vb_eps}
\end{align}


\vspace{0.5cm}
{\bf entanglement velocity.} 
We use the recursion relation for the purity from Ref.~\cite{nahum_quantum_2017}, adapted to the case with gate dilution: if $\mathcal{P}(x,t)$ is the subsystem purity for a cut at bond $x$ at time $t$, we have
\begin{equation}
    \mathcal{P}(x,t) = \left\{
    \begin{aligned} 
    & (1-\epsilon) \mathcal{P}(x,t-1) + \epsilon \frac{q}{q^2+1}\left[ \mathcal{P}(x-1,t-1) + \mathcal{P}(x+1,t-1)\right] 
     \text{ if } x+t \text{ is even,}    \\
    & \mathcal{P}(x,t-1) \text{ if } x+t \text{ is odd.}    
    \end{aligned} \right.
\end{equation}
Assuming $\mathcal{P}(x,t)\sim q^{-v_E t}$, we obtain a quadratic equation for $q^{v_E}$:
\begin{equation}
    1 = (1-\epsilon)q^{2v_E} + \epsilon \frac{2q}{q^2+1} q^{v_E}
\end{equation}
which can be solved to obtain
\begin{equation}
    v_E(\epsilon) = \log_q \left[\frac{\epsilon}{2(1-\epsilon)} \frac{2q}{q^2+1}
                \left( \sqrt{1+ \frac{1-\epsilon}{\epsilon^2} \left(\frac{q^2+1}{q}\right)^2} -1 \right) \right].
                \label{eq:supp_ve_eps}
\end{equation}

We note that the recursion for $\mathcal{P}(x,t)$ can be thought of as a random walk with steps 
\begin{equation} 
\left\{ \begin{aligned}
(x,t)\mapsto (x+1,t-1) &\ [\text{prob. } p_+ = \epsilon q/(q^2+1)] \\
(x,t)\mapsto (x+1,t-2) &\ [\text{prob. } p_0 = 1-\epsilon] \\
(x,t)\mapsto (x-1,t-1) &\ [\text{prob. } p_- = \epsilon q/(q^2+1)]
\end{aligned} \right. \label{eq:supp_purity_rw}
\end{equation}
where the ``probabilities'' $p_\pm$, $p_0$ are not normalized: $p_+ + p_0 + p_- <1$. The purity at time $t$ is given by the total survival probability. This point of view will be useful for connecting $v_E$ to $\gamma$ later.


\vspace{0.5cm}
{\bf Saddle-point butterfly velocity.} 
We simulate the random walk for the Pauli endpoint, Eq.~\eqref{eq:supp_vb_rw}, starting from an initial probability distribution $p_0(x) = \delta_{x,x_0}$ and obtain a late-time distribution $p_t(x)$. Then we fit $\overline{q^{-x(t)}} = \sum_x p_t(x) q^{-x}$ to $q^{-v_B^{\rm sp} t}$, following the discussion in Sec.~\ref{sec:supp_meanfield}.

We note that this calculation is formally identical to the one used for computing the entanglement velocity $v_E$ from an operator-spreading picture in Ref.~\cite{von_keyserlingk_operator_2018} (in particular see Eq. (24) therein). 
There, the purity of a subsystem $A$ is computed by analyzing the propagation of operator endpoints across the subsystem boundary; an exponential weighting factor $q^{-x}$ arises from the counting of (diagonal) Pauli operators whose endpoint is initially a distance $x$ from the entanglement cut, giving rise to the same sum.
It follows that $v_B^{\rm sp} = v_E$.

\begin{figure}
    \centering
    \includegraphics[width=\textwidth]{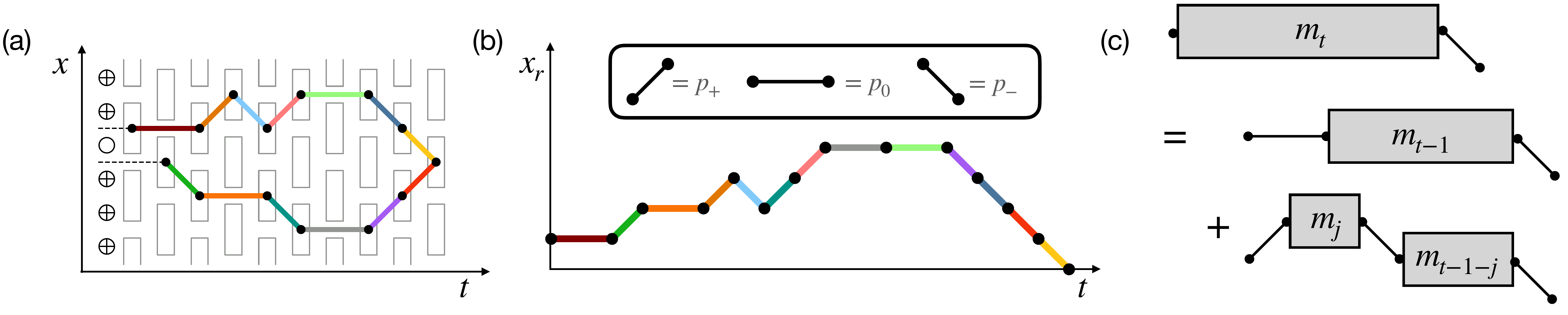}
    \caption{Random walk picture for the relaxation of Pauli density with gate dilution ($\epsilon < 1$).
    (a) Random walks for the two endpoints $x_{1,2}$ of the $\vacuum$ domain.
    Rectangles represent gates in the brickwork circuit; walkers may only hop when on a gate.
    (b) Generalized Motzkin walk for the relative coordinate $x_r = x_2-x_1$. Steps are color-coded to match (a). The three possible steps ($\delta x_r = \pm 1, 0$) have probabilities $p_{\pm}$, $p_0$. 
    (c) Diagrammatic representation of the recursion relation Eq.~\eqref{eq:supp_motzkin_recursion} for the first-return probability (summation over $j=0,\dots t-1$ is implied). }
    \label{fig:motzkin}
\end{figure}

\vspace{0.5cm}
{\bf Pauli density relaxation rate}.
As explained in Sec.~\ref{sec:supp_rw}, the relaxation of $\overline{n_i}(t)$ is determined by the annihilation probability of two biased random walkers $x_1 \leq x_2$ (representing the boundaries of a domain of $\vacuum$ in a background of $\free$ states). Without dilution it is straightforward to derive $\gamma = 2\ln\frac{q^2+1}{2q}$ (equal to $2\ln(q) v_E$) by considering the random walk of the relative coordinate.
The $\epsilon < 1$ case can be analyzed in a similar way. 
Each walker $x_{1,2}$ lives on a bond on a 1D lattice, and can hop only when a gate acts on that bond. It is helpful to consider a 2D square lattice in space-time with sites corresponding to gates in the brickwork circuit, i.e. $(x,t)$ with $x+t$ even, Fig.~\ref{fig:motzkin}(a). 
The walkers start from $(x_1,t) = (0,1)$ and $(x_2,t) = (1,0)$ and can hop as follows:
\begin{equation}
    (x_1,t)\mapsto \left\{ 
    \begin{aligned} 
    (x_1-1,t-1) & \ [\text{prob. } p_+ = \epsilon (1-a)] \\
    (x_1,t-2) & \ [\text{prob. } p_0 = 1- \epsilon] \\
    (x_1+1, t-1) & \ [\text{prob. } p_- = \epsilon a]
    \end{aligned} \right.
    \qquad 
    (x_2,t)\mapsto \left\{ 
    \begin{aligned} 
    (x_2-1,t-1) & \ [\text{prob. } p_- = \epsilon a] \\
    (x_2,t-2) & \ [\text{prob. } p_0 = 1- \epsilon] \\
    (x_2+1, t-1) & \ [\text{prob. } p_+ = \epsilon (1-a)]
    \end{aligned} \right.
\end{equation}
The probability that they annihilate (i.e. cross for the first time) at time $\tau$ is given by a sum over loops as in Fig.~\ref{fig:motzkin}(a). It is easy to see that such loops are in one-to-one correspondence with a type of generalized {\it Motzkin walks} for the relative coordinate $x_r \equiv x_2-x_1$, where the flat step ($\delta x_r=0$) covers two time steps ($\delta t=2$) unlike the conventional Motzkin walks where all three possible steps have  $\delta t = 1$. One such walk is shown in Fig.~\ref{fig:motzkin}(b); it describes the evolution of the relative coordinate in Fig.~\ref{fig:motzkin}(a).

We denote the probability of annihilation at time $2t+1$ by $p_- m_t$ (it is convenient to factor out the last step, which is necessarily $\delta x_r = -1$, hence the $p_-$). Then, we have $m_0 = 1$ and the Motzkin-like recursion relation 
\begin{equation}
m_t = p_0 m_{t-1} + p_+ p_- \sum_{j=0}^{t-1} m_j m_{t-1-j}, \label{eq:supp_motzkin_recursion}
\end{equation}
The idea is that the first step in the walk can either be $\delta x_r =0$, or $\delta x_r = +1$.
The former gives $p_0m_{t-1}$ (as after the flat step, the problem remains unchanged up to $t\mapsto t-1$). 
In the latter case, the walk must at some point return to $x_r=1$ ({\it en route} to $x_r=0$); letting $2(j+1)$ be the time of first return to $x_r=1$ (note this must be even), we obtain a term $p_+ m_j p_- m_{t-1-j}$. This is illustrated diagrammatically in Fig.~\ref{fig:motzkin}(c).
Finally one must sum over all possible values of $j$, obtaining Eq.~\eqref{eq:supp_motzkin_recursion}.
We solve the recursion numerically and extract the scale $\gamma(\epsilon)$ from $e^{-\gamma(\epsilon)t} \sim m_t$.

We note that the result depends on $p_+$ and $p_-$ only through their product $p_+p_- = [\epsilon q / (q^2+1)]^2$.
The purity calculation, Eq.~\eqref{eq:supp_purity_rw}, has $p_+ = p_- = \epsilon q/(q^2+1)$, giving the same value for the product $p_+ p_-$.
It follows that the return probability is the same in the two cases. 
Furthermore, in the purity calculation the random walk is unbiased ($p_+ = p_-$), so that up to a power-law correction the return probability is the same as the total survival probability; the latter is what maps onto the purity, $\sim q^{-v_E t}$.
Adjusting for the fact that the computation of $\gamma$ involves two random walks (for the left and right endpoints on the $\vacuum$ domain, Fig.~\ref{fig:motzkin}(a)), or equivalently one Motzkin walk of doubled length (Fig.~\ref{fig:motzkin}(b)), we obtain $e^{-\gamma t} \sim q^{-2v_Et}$ and thus $\gamma = 2\ln(q) v_E$.


\section{Non-contiguous operators \label{sec:supp_noncontiguous}}

Here we analyze the performance of shallow shadows for learning non-contiguous operators. 
We focus on two cases of interest: Pauli operators with a sufficiently low, finite density of ``holes'' (i.e. identity operators), and Pauli operators whose support is made of multiple widely-separated connected components. 

\subsection{Low density of holes \label{sec:supp_lowdensity}}

Let us consider a contiguous segment $A$ of length $\ell$, and a Pauli operator $O_B$ supported on a non-contiguous subsystem $B\subseteq A$ with $|B| = k < \ell$, as sketched in Fig.~\ref{fig:supp_noncontig}(a). Let us further assume that $k/\ell$ is sufficiently large and that the support $B$ does not have large holes; e.g., one can take $O_B$ to be a typical random Pauli operator (with identity matrices allowed) on the $A$ segment. 
At depth $t = t^\star(\ell) = \gamma^{-1} \ln(\ell) + \dots $, we expect the weight distribution of the twirled operator $O_B(t)$ to look essentially indistinguishable from that of another operator $O'_A(t)$ that had no holes at $t=0$.
That is to say, we expect the interior of both $O'_A(t)$ and $O_B(t)$ to have nearly equilibrated by time $t^\star(\ell)$, thus erasing memory of the initial conditions in the weigth distributions.
Therefore we expect the late-time scaling $\shadownorm{O_B}^2 \simeq q^{|B|} e^{\gamma t}$, which at $t = t^\star(\ell)$ yields
\begin{equation}
    \shadownorm{O_B}^2 \simeq \ell q^\ell.
\end{equation}
The shadow norm for random Pauli measurements ($t = 0$), for comparison, is $\shadownorm{O_B}^2 = (q+1)^k$. 
Therefore evolving to depth $t^\star(\ell)>0$ is advantageous if and only if
\begin{equation}
    \frac{k}{\ell} \geq n_{\rm sp} + \frac{\ln \ell}{\ell \ln(q+1)},
    \qquad
    n_{\rm sp} = \frac{\ln(q)}{\ln(q+1)}.
\end{equation}
Here $n_{\rm sp}$ is the ``saddle point density'' already encountered in Sec.~\ref{sec:supp_velocities}.
For qubits ($q = 2$) this takes the value $n_{\rm sp} = 1/ \log_2(3) \simeq 0.631 $, which is notably less than the average density of a random Pauli operator, $3/4 = 0.75$. 
The same is true at large $q$, where we have $n_{\rm sp} = 1 - 1/q  + O(1/q^2)$ whereas the average density of a random Pauli operator is $1-1/q^2$. 

Thus twirling to depth $t^\star$ is advantageous not only for operators with contiguous support, but much more generally for operators with a sufficiently low density of holes. The threshold Pauli density, $k/\ell \geq n_{\rm sp} = \ln(q) / \ln(q+1)$ (at large $\ell$), is low enough that this includes typical random Pauli operators drawn from a given length-$\ell$ segment.
Such operators, despite having the same average density as an equilibrated operator ($1-q^{-2}$), are not at equilibrium: 
their weight distribution is a $\delta$-function, $\pi(w) = \delta_{w, (1-q^{-2})\ell }$, as opposed to a binomial distribution of the same mean. Under local twirling we thus have the shadow norm
\begin{equation}
    \shadownorm{O}^2 = (q+1)^k = (q+1)^{(1-q^{-2})\ell},
\end{equation}
e.g. for qubits this yields $\simeq 2.28^\ell$ (it has the same form as the upper bound $(q+1)^{\overline{w}}$, cf Sec.~\ref{sec:supp_ub}), compared to the scaling $\sim \ell 2^\ell$ achieved at the optimal depth. 

\subsection{Multiple widely-separated components \label{sec:supp_multipoint}}

\begin{figure}
    \centering
    \includegraphics[width=0.9\textwidth]{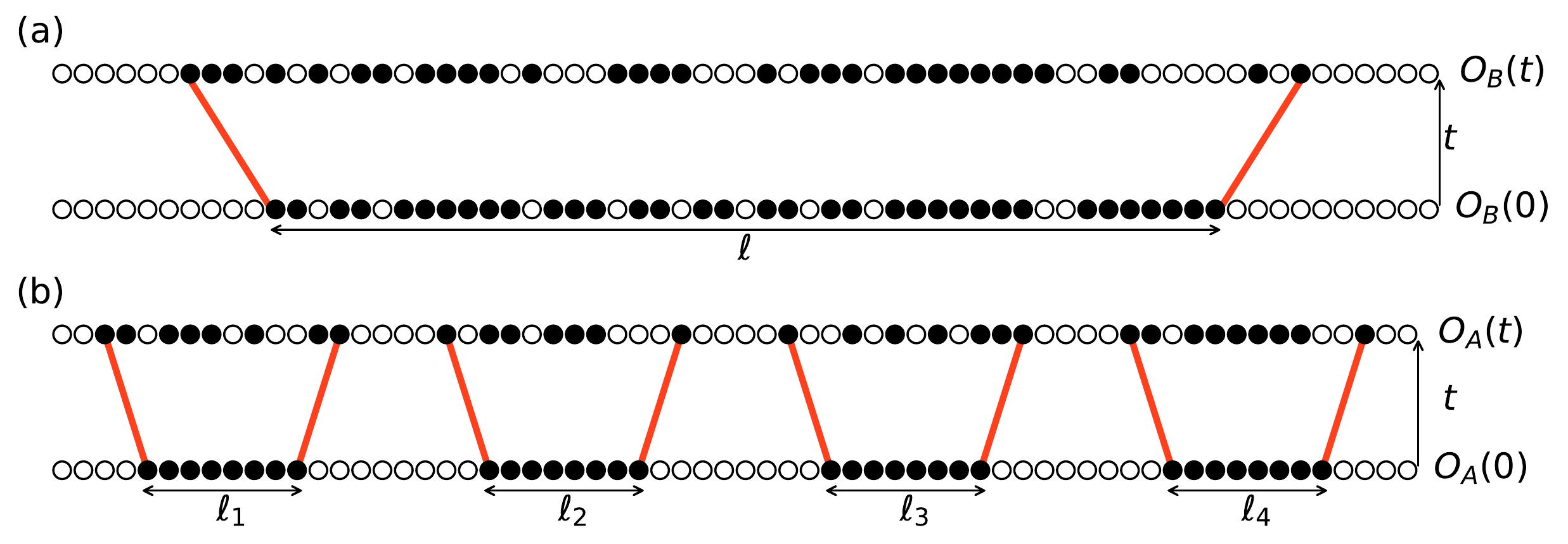}
    \caption{Effect of finite-depth twirling on non-contiguous Pauli operators. 
    (a) A Pauli operator $O_B$ supported on a non-contiguous subset $B$ of a length-$\ell$ segment, as discussed in Sec.~\ref{sec:supp_lowdensity}. After twirling, $O_B(t)$ is in general a superposition of many Pauli operators with a near-equilibrium weight distribution; the shadow norm is dominated by operators with Pauli density near the ``saddle-point'' value $n_{\rm sp} = \ln(q) / \ln(q+1)$. One such operator is shown. Whenever the Pauli density of the initial operator (number of $\occupied$ sites out of $\ell$) is larger than $n_{\rm sp}$, shallow shadows can be more sample-efficient than local twirling.
    (b) A tensor product of widely-separated, contiguous Pauli operators, $O_A = \bigotimes_i O_{A_i}$, supported on segments $A_i$ of lengths $\ell_i$, as discussed in Sec.~\ref{sec:supp_multipoint}. If the light cones don't overlap (as shown), then the shadow norm factors across components. Shallow shadows are more sample-efficient than local twirling for any finite number of segments (asymptotically in large $\ell$), and also for diverging number of segments as long as each $\ell_i$ is sufficiently large. 
    }
    \label{fig:supp_noncontig}
\end{figure}

Another situation of interest is that of a Pauli operator $O_A$ supported on a set $A = \bigcup_{i=1}^n A_i$ which is the union of $n>1$ disconnected, widely-separated segments $A_i$, $i=1,\dots n$. We focus on the case where each $A_i$ is contiguous (no interior holes), though this could be relaxed to density $> n_{\rm sp}$ using the result above. The situation, sketched in Fig.~\ref{fig:supp_noncontig}(b), might arise e.g. when trying to learn {\it multi-point correlation functions} of a given string-like operator. 

We take the length of each segment to be $\ell_i = |A_i|$, set $\ell_{\rm max} = \max_i \ell_i$, and consider twirling to a depth $t = t^\star(\ell_{\rm max})  = \gamma^{-1} \ln(\ell_{\rm max}) + \dots$. Further, we take any two segments $A_i$, $A_j$ to be separated by a distance $\gg \ln(\ell_{\rm max})$, so that on the time scales of interest the light cones emanating from each segment do not overlap, and the twirled operator $O_A(t)$ remains a tensor product $\bigotimes_i O_{A_i}(t)$. 
It is straightforward to see that the weigth distribution factors in this case: $\pi_{A,t}(w) = \sum_{w_1,\dots w_n} \delta_{w = w_1+\dots +w_n} \prod_i \pi_{A_i,t}(w_i)$ with $w = w_1 + \dots + w_n$ being the decomposition of the total weight into weights for each segment. This implies
\begin{equation}
    \shadownorm{O_A}^{-2} = \sum_w \pi_{A,t}(w) (q+1)^{-w} 
    = \prod_{i=1}^n  \sum_{w_i} \pi_{A_i,t}(w_i) (q+1)^{-w_i}
    = \prod_{i=1}^n \shadownorm{O_{A_i}}^{-2} .
\end{equation}

At depth $t^\star(\ell_{\rm max})$, we have $\shadownorm{O_{A_i}}^2 \simeq q^{\ell_i} e^{\gamma t^\star} = \ell_{\rm max} q^{\ell_i} $, and thus
\begin{equation}
    \shadownorm{O_A}^2 \simeq \ell_{\rm max}^n q^k
\end{equation}
using the fact that by definition $k = \sum_{i=1}^n \ell_i$. 
For any finite $n$, this gives ${\rm poly}(k) q^k$ which is asymptotically advantageous over the $t = 0$ scaling $(q+1)^k$.
Thus shallow shadows provide a sampling advantage also for $n$-point functions of large contiguous operators, for any finite $n$. 

Finally, we may also take finite segments of constant, sufficiently large size $\ell_i\equiv \ell$, and take their number $n$ to go to infinity proportionally with $k$ ($n = k/\ell$). This gives a shadow norm scaling as $\ell^{k/\ell} q^k = (q \ell^{1/\ell})^k$, which is also lower than $(q+1)^k$ as long as $\ell$ is large enough. 
This illustrates the fact that shallow shadows are advantageous over the $t = 0$ protocol whenever the operator support $A$ has a volume $|A|$ that is sufficiently larger than its boundary $|\partial A|$.
Surprisingly we do not need a parametric separation; in this case ($|A| = k$ and $|\partial A| = 2n$) a sufficienlty large but finite ratio ($\ell = k/n$) is enough to obtain a sampling advantage.

\section{Details on iMPS numerical simulations}

Here we present details of the compuational method for simulating the dynamics of $p_{\mathbf n}$ under the averaged twirling circuit.

We use an infinite matrix product state (iMPS) description of $p_{\mathbf n}$. Due to the brickwork structure of the twirling circuits, this iMPS has a unit cell of two sites; thus 
\begin{equation}
p_{\mathbf n} = \sum_{\boldsymbol \alpha } \prod_i A_{\alpha_i, \alpha_{i+1}}^{n_i} B_{\alpha_{i+1}, \alpha_{i+2}}^{n_{i+1}}
\end{equation}
in terms of tensors $A$, $B$ of dimension $(2, \chi_1, \chi_2)$ and $(2,\chi_2,\chi_1)$ respectively, with $\chi_{1,2}\leq \chi$ for some fixed maximum bond dimension $\chi$.

In order to obtain the shadow norm of a length-$k$ fully packed Pauli string, we could start with an intial state $p_{\mathbf n}^{\rm init} = \prod_{i\in R} \delta_{n_i,1} \prod_{i\notin R} \delta_{n_i,0}$, evolve it over $t$ steps, contract it with the final boundary condition $f_{\mathbf n} = (q+1)^{-\sum_i n_i}$, and repeat for each value of $k$.
A more efficient alternative is to instead start with the final boundary condition $f_{\mathbf n}$, which is trivially expressible as an iMPS with bond dimension $\chi = 1$ and tensors $A^n_{\alpha\beta} = (q+1)^{-n}$ (with dummy indices $\alpha = \beta = 1$) and $B = A$; evolve the iMPS ``backwards'' for $t$ steps, i.e. under $\mathbb{M}_t^T$; and then compute the shadow norm for all values of $k$ at once.

The evolution of the iMPS proceeds as follows.
Under even layers of the twirling circuits, tensors $A$ and $B$ are merged by the action of an averaged gate $M = (1-\epsilon)I + \epsilon M_{\rm Haar}$, as in Eq.~\eqref{eq:supp_M_haar}.
The tensor contraction reads
\begin{equation}
C^{n_i,n_{i+1}}_{\alpha_i, \alpha_{i+2}} = \sum_{m_i, m_{i+1}, \alpha_{i+1}} (M^T)_{m_i,m_{i+1}}^{n_i, n_{i+1}} A^{m_i}_{\alpha_i,\alpha_{i+1}} B^{m_{i+1}}_{\alpha_{i+1},\alpha_{i+2}}
\end{equation}
We then split $C$ into two tensors $A'$ and $B'$ of bond dimension at most $\chi$ in the standard way (by performing a singular-value decomposition and keeping the largest $2\chi$ values).
This defines an update procedure $A, B \mapsfrom f(A,B)$.
Under odd layers of the twirling circuit, we simply have $B, A \mapsfrom f(B,A)$. 
We iterate these two steps a total of $t$ times ($t/2$ each).

Finally, to compute the shadow norm of a Pauli string of size $k$, we must contract the backwards-evolved iMPS with an initial condition $|\cdots \vacuum \vacuum \overbrace{\occupied\occupied\cdots\occupied\occupied}^k \vacuum \vacuum\cdots)$.
To this end, we compute the left and right ``environments'' by finding the leading left- and right-eigenvectors of the transfer matrix $\mathbb{T}_{\alpha\beta}^{\vacuum} \equiv \sum_\kappa A^0_{\alpha\kappa} B^0_{\kappa\beta}$. $\mathbb{T}^{\vacuum}_{\alpha\beta}$ has a unique unit eigenvalue (associated to the conservation of total probability under the stochastic dynamics); we label the corresponding left- and right-eigenvectors $E_l$, $E_r$.
The shadow norm is then given by
\begin{equation}
    \shadownorm{O_A}^2 = \bra{E_l} \left( \mathbb{T}^{\occupied} \right)^{k/2} \ket{E_r}
\end{equation}
where $\mathbb{T}^{\occupied}_{\alpha\beta} = A^1_{\alpha\kappa} B^1_{\kappa\beta}$ and we assumed $k$ even for simplicity. 
Note that the results for all values of $k$ can be computed in one ``sweep'', by computing the enviroments $E_{l,r}$ once and diagonalizing the transfer matrix $A^1 B^1$.


\section{Results on Brownian circuit}

\begin{figure}[t]
    \centering
    \includegraphics[width=.4\textwidth]{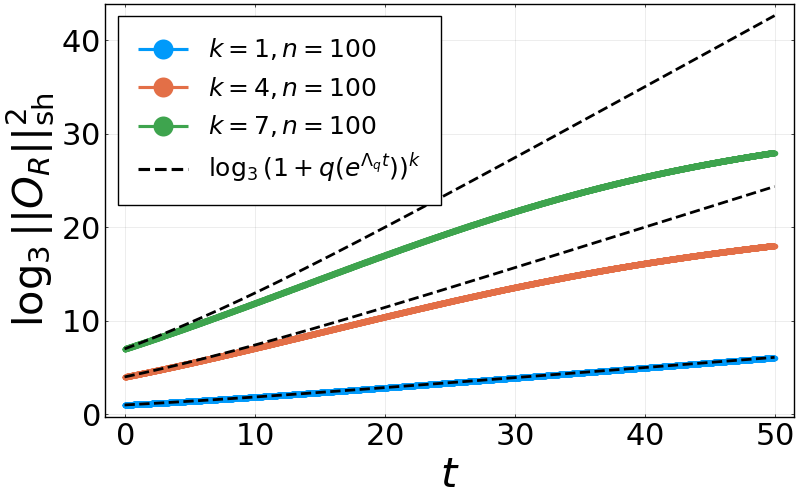}
    \hspace{.5in}
    \includegraphics[width=.4\textwidth]{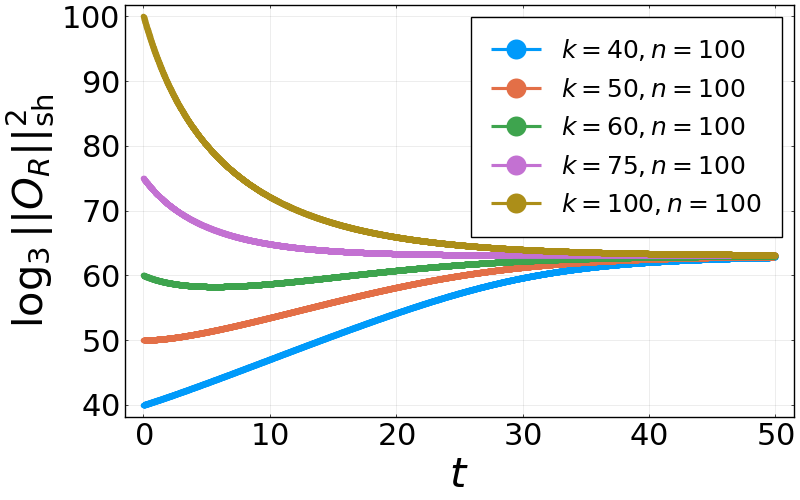}
    \caption{Results for the shadow norm in the Brownian circuit model, from numerical integration of Eq.~\eqref{eq:pi_w_brownian_master_equation} for a system of $n=100$ qubits. 
    (a) Few-body Pauli operators, $k\ll n$: the optimal depth is $t^\star = 0$.
    (b) Large Pauli operators: the optimal depth becomes nonzero when the initial density $k/n$ exceeds a finite fraction.}
    \label{fig:brownian}
\end{figure}

Here we consider the time evolution of the weight distribution function (Eq.~\eqref{eq:weight_dist} of main text) under a ``Brownian circuit'', see for example Refs.~\cite{lashkari_towards_2013, zhou_operator_2019, piroli_random_2020} for introductions to such models.
The system we consider has $n$ qudits undergoing random pairwise untiary gates.
Specifically, within each ``update step'' with probability $\epsilon$ we apply a random unitary gate on a pair of randomly chosen qudits, and with probability $1-\epsilon$ we do nothing.
Here the parameter $\epsilon$ is the same dilution constant introduced in the main text for 1d circuits.
A ``time step'' consists of $n$ such update steps, to be comparable with finite dimensional circuits.

Within this model, the weight distribution function $\pi_w(t)$ evolves under the following master equation,
\begin{align}
\label{eq:pi_w_brownian_master_equation}
    \frac{d \pi_w(t)}{dt} = \frac{n \epsilon}{\binom{n}{2}}
    \left\{
        \frac{(q^2-1)^2}{q^4-1}
        \left[
        (w-1)(n-w+1) \pi_{w-1}
        - w(n-w) \pi_{w}
        \right]
        +
        \frac{2(q^2-1)}{q^4-1}
        \left[
        \binom{w+1}{2} \pi_{w+1}
        - \binom{w}{2} \pi_{w}
        \right]
    \right\}.
\end{align}
The operator weight can either increase or decrease under random unitary gates, as captured by the first and second term, respectively.
For the relaxation of a single Pauli operator of weight $k$, as considered in the main text, we set the initial condition $\pi_w = \delta_{w, k}$.

We first focus on the early time dynamics with circuit depth at most $O(k)$, and for operator weight $k \ll n$.
The equation is greatly simplified in this regime, as it suffices to focus on $\pi_w$ with $w \ll n$,
\begin{align}
    \frac{d \pi_w(t)}{dt} = 
        \Lambda_q \left[
        (w-1) \pi_{w-1}
        - w \pi_{w}
    \right], \text{ where } \Lambda_q = \frac{(2 \epsilon)(q^2-1)}{q^2+1}.
\end{align}
The generating function of $\pi_w(t)$, defined as $f(z,t) = \sum_{w=0}^\infty \pi_w(t) z^w$, evolves under the following partial differential equation
\begin{align}
    \frac{\partial f(z,t)}{\partial t} = \Lambda_q z(z-1) \frac{\partial f(z,t)}{\partial z}, \text{ with initial condition } f(z,t=0) = z^k,
\end{align}
from which we obtain 
\begin{align}
    f(z,t) = \left( \frac{z e^{-\Lambda_q t}}{1-z(1-e^{-\Lambda_q t})} \right)^k.
\end{align}
Finally, comparing with Eq.~\eqref{eq:lambda_vs_weight}, we can immediately read off $\lambda_A$ from $f(z,t)$,
\begin{align}
    \lambda_A = f(z = (1+q)^{-1}, t) = \left(\frac{1}{1+q e^{\Lambda_q t}}\right)^k,
\end{align}
which is a monotonically decreasing function.
In Fig.~\ref{fig:brownian}(a) we plot $\left\Vert O_A \right\Vert_{\rm sh, avg} = \lambda_A^{-1}$ for $k \ll n$, from numeical solutions of Eq.~\eqref{eq:pi_w_brownian_master_equation} at $q=2$, and find good agreement.
The optimal circuit depth is thus at $t^\star=0$, as consistent with the absence of comparable boundary and bulk effects.

On the other hand, when $k = O(n)$, $\lambda_A$ can exhibit different behaviors depending on the operator density $k/n$, as we observe in Fig.~\ref{fig:brownian}(b).
When $k/n < 1/2$, the boundary operator growth  always dominates over the bulk operator relaxation, resulting in a monotonically increasing $\left\Vert O_A \right\Vert_{\rm sh, avg}$, much like the case $k \ll n$.
When $k/n > 1-1/q^2$, the operator density is above its equilibrium value, and the bulk relaxation process always dominates. 
When $1/2 \lesssim k/n \lesssim 1-1/q^2$,
$\lambda_A$ is non-monotonic, where the two effects are comparable at early times.

\end{document}